\documentclass{vldb}

\usepackage{graphicx,epsfig,color,endnotes,alltt}
\usepackage{subfigure}
\usepackage{url}
\usepackage{balance}
\usepackage{algorithm}
\usepackage[noend]{algorithmic}
\usepackage{subfigure}
\usepackage{fancyvrb}
\usepackage[noindent, nolineno]{lgrind}
\usepackage{listings}
\usepackage{epstopdf}
\usepackage{array}
\usepackage{multirow}
\usepackage{amsmath}
\usepackage{bm}
\usepackage{slashbox}

\newcommand{\saven}[1]{{{#1}}}

\newcommand{\hejiong}[1]{{#1}}

\newcommand{\hj}[1]{{{#1}}}

\newcommand{\tosaven}[1]{{#1}}

\begin{document}

\title{Revisiting Co-Processing for Hash Joins on the Coupled CPU-GPU Architecture}

\numberofauthors{3}

\author{
\alignauthor
Jiong He\\
       \affaddr{Nanyang Technological University}\\
\alignauthor
Mian Lu\\
       \affaddr{A*STAR Institute of HPC, Singapore}\\
\alignauthor
Bingsheng He\\
       \affaddr{Nanyang Technological University}\\
}

\maketitle
\begin{abstract}

Query co-processing on graphics processors (GPUs) has become an
effective means to improve the performance of main memory databases.
However, the relatively low bandwidth and high latency of the PCI-e
bus \hj{are usually bottleneck issues} for co-processing. Recently,
\emph{coupled} CPU-GPU architectures have received a lot of
\hejiong{attention}, e.g. AMD APUs with the CPU and the GPU
integrated into a single chip. That opens up new opportunities for
optimizing query co-processing. In this paper, we experimentally
revisit hash joins, one of the most important join algorithms for
main memory databases, on a coupled CPU-GPU architecture.
Particularly, we study the \emph{fine-grained} co-processing
mechanisms on hash joins with and without partitioning. The
co-processing outlines \hj{an interesting design space}. We extend
existing cost models to automatically guide decisions on the design
space. Our experimental results on a recent AMD APU show that (1)
the coupled architecture enables fine-grained co-processing and
cache reuses, which are inefficient on \emph{discrete} CPU-GPU
architectures; (2) the cost model can automatically guide the design
and tuning knobs in the design space; (3) fine-grained co-processing
achieves up to 53\%, 35\% and 28\% performance improvement over
CPU-only, GPU-only and conventional CPU-GPU co-processing,
respectively. \saven{We believe that the insights and implications
from this study are initial yet important for further research on
query co-processing on coupled CPU-GPU architectures.}

\end{abstract}

\section{Introduction} \label{sec:introduction}

Query co-processing on GPUs (or Graphics Processing Units) have been
an effective means to improve the performance of main memory
databases(e.g.,~\cite{RelationalQueryCoProcessing,wenbinVLDB,databaseOperatorsOnGPU,hashjoinOnCGwithPCIeconstraint,sortHashRevisited,GPUJoinRevisited}).
Compared with CPUs, GPUs have rather high memory bandwidth and
massive thread parallelism, which is ideal for data parallelism in
query processing. Designed as a co-processor, the GPU is usually
connected to the CPU with a PCI-e bus. So far, most research studies
on GPU query co-processing have been conducted on such
\emph{discrete} CPU-GPU architectures. The relatively low bandwidth
and high latency of the PCI-e bus are usually bottleneck
issues~\cite{RelationalQueryCoProcessing,wenbinVLDB,hashjoinOnCGwithPCIeconstraint,GPUJoinRevisited}.
GPU co-processing algorithms have to be carefully designed so that
the overhead of data transfer on PCI-e is minimized. Recently,
hardware vendors have attempted to resolve this overhead with new
architectures. We have seen integrated chips with \emph{coupled}
CPU-GPU architectures. The CPU and the GPU are integrated into a
single chip, avoiding the costly data transfer via \hj{the }PCI-e
bus. For example, the AMD APU (Accelerated Processing Unit)
architecture integrates the CPU and the GPU into a single chip, and
Intel released their latest generation Ivy Bridge processor in late
April, 2012. \hejiong{These} new heterogeneous architectures
potentially open up new optimization opportunities for GPU query
co-processing. Since hash joins have been extensively studied on
both CPUs and GPUs and are one of the most important join algorithms
for main memory databases, this paper revisits co-processing for
hash joins on the coupled CPU-GPU architecture.

Before exploring the new opportunities by the coupled architecture,
let us analyze the performance issues of query co-processing on
discrete architectures. A number of hash join algorithms have been
developed for CPU-GPU co-processing on the discrete
architecture~\cite{databaseOperatorsOnGPU,GPUJoinRevisited,sortHashRevisited}.
The data transfer on the PCI-e bus is an inevitable overhead,
\hejiong{although it is feasible to reduce this overhead} by
overlapping computation and data transfer. \hejiong{This data
transfer} leads to \emph{coarse-grained} query co-processing
schemes. Most studies off-load the entire join to the GPU, and the
CPU is usually under-utilized during the join process. On the other
hand, co-processing on the discrete architecture under-utilizes the
data cache, because the CPU and the GPU have their own caches. This
prohibits cache reuse among processors to reduce the memory stall,
which is one of the major performance factors for hash
joins~\cite{Ailamaki1999, Manegold2000}.




By integrating the CPU and the GPU into the same chip, the coupled
architecture eliminates the data transfer overhead of \hejiong{the}
PCI-e bus. Existing query co-processing algorithms
(e.g.,~\cite{GPUJoinRevisited,sortHashRevisited}) can
\hejiong{immediately} gain this performance improvement. However, a
natural question is whether and how we can further improve the
performance of co-processing on the coupled architecture.
Considering the above-mentioned performance issues of co-processing
on discrete architectures, we find that \hejiong{the} two hardware
features of the \hejiong{coupled} architecture lead to significant
implications on the effectiveness of co-processing.

\hejiong{Firstly}, unlike discrete architectures \hejiong{where} the
CPU and the GPU have their own memory hierarchy, the CPU and the GPU
in the coupled architecture share the same physical main memory.
Furthermore, they can share some levels of the cache hierarchy
(e.g., the CPU and the GPU in the AMD APU share \hejiong{the} L2
data cache). The CPU and the GPU can communicate at the memory (or
even cache) speed, and more \emph{fine-grained} co-processing
designs and data sharing are feasible, which \hj{are} inefficient on the
discrete architecture.

\hejiong{Secondly}, due to the limited chip area, the GPU in the coupled
architecture is usually less powerful than the high-end GPU in
discrete architectures (see Table~\ref{table:APUconfiguration} of
Section~\ref{sec:related}). The GPU in the coupled architecture
usually cannot deliver \hejiong{a} dominant performance speedup as it does in
discrete architectures. Therefore, a good co-processing scheme must
keep both processors busy as well as carefully \hejiong{assigning} workloads to
them for the maximized speedup.

%
%

In this paper, we revisit co-processing schemes for hash joins on
such CPU-GPU coupled architectures. Specifically, we examine both
hash joins with and without partitioning, and explore \hejiong{the} design
space of fine-grained co-processing on the coupled architecture.
Additionally, we consider a series of design tradeoffs such as
whether the hash table should be shared or separated between the two
processors as well as different co-processing granularities. We
extend existing cost models to predict the performance of hash joins
with different co-processing schemes, and thus guide the decisions
on co-processing design space.


We implement all co-processing schemes of hash joins with OpenCL on
AMD APU A8 3870K. The major experimental results are summarized as
follows.

First, we evaluate the co-processing schemes on the (emulated)
discrete architecture in comparison with the coupled architecture,
and show that (1) conventional co-processing of hash
joins~\cite{databaseOperatorsOnGPU, sortHashRevisited} can achieve
only marginal performance improvement on the coupled architecture;
(2) the coupled architecture enables fine-grained co-processing and
cache reuse optimizations, which are inefficient/infeasible on
discrete architectures (Section~\ref{subsec:discreteeval}).

Second, we evaluate the cost model and show that our cost model is
able to effectively guide the decision on the optimizations in the
design space (Section~\ref{subsec:costmodel}).

Third, we evaluate a number of design tradeoffs on the fine-grained
co-processing, which are important for the hash join performance
(Section~\ref{subsec:tradeoff}).

Fourth, we evaluate a number of co-processing variants, and show
that fine-grained co-processing achieves up to 53\%, 35\% \hj{and 28\%}
performance improvement compared with CPU-only, GPU-only and
conventional CPU-GPU co-processing, respectively
(Section~\ref{subsec:cop}).

To the best of our knowledge, this is the first systematic study of
hash join co-processing on the coupled architecture. \saven{We
believe that the insights and implications from this study can shed
light on further research of query co-processing on CPU-GPU coupled
architectures.}



{\bf Organization.} The remainder of this paper is organized as
follows. In Section~\ref{sec:related}, we briefly introduce
background and related work. In Section~\ref{sec:design}, we revisit
fine-grained CPU-GPU co-processing schemes for hash joins. The cost
model is described in Section~\ref{sec:model}. We present the
experimental results in Section~\ref{sec:evaluation} and conclude in
Section~\ref{sec:conclusion}.

\section{Preliminaries and Related Work} \label{sec:related}

We first introduce CPU-GPU architectures and OpenCL, and then review
the related work on hash joins.

\subsection{Coupled CPU-GPU Architecture}

GPUs have \hejiong{emerged as} promising hardware co-processors to
speedup various applications, such as scientific computing
\cite{GPGPU:DGEMMOnGPU} and database operations
\cite{databaseOperatorsOnGPU}. With massive thread parallelism and
high memory bandwidth, GPUs are suitable for applications with
massive data parallelism and high
computation intensity. 



\begin{figure}
\centerline{
\subfigure[Discrete]{\includegraphics[width=0.22\textwidth]{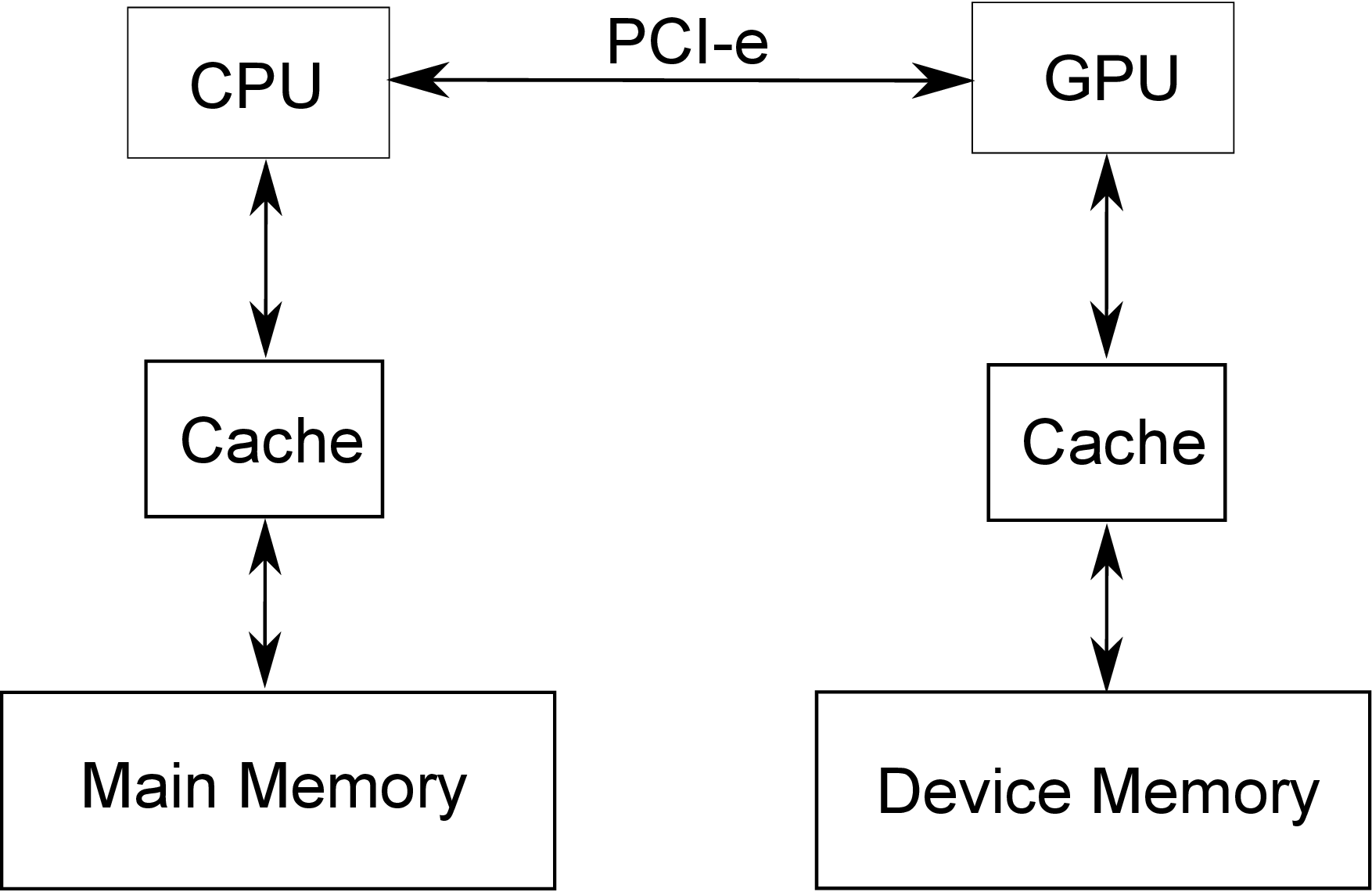}
\label{fig:generalDiscreteSystem}} \hfil
\subfigure[Coupled]{\includegraphics[width=0.22\textwidth]{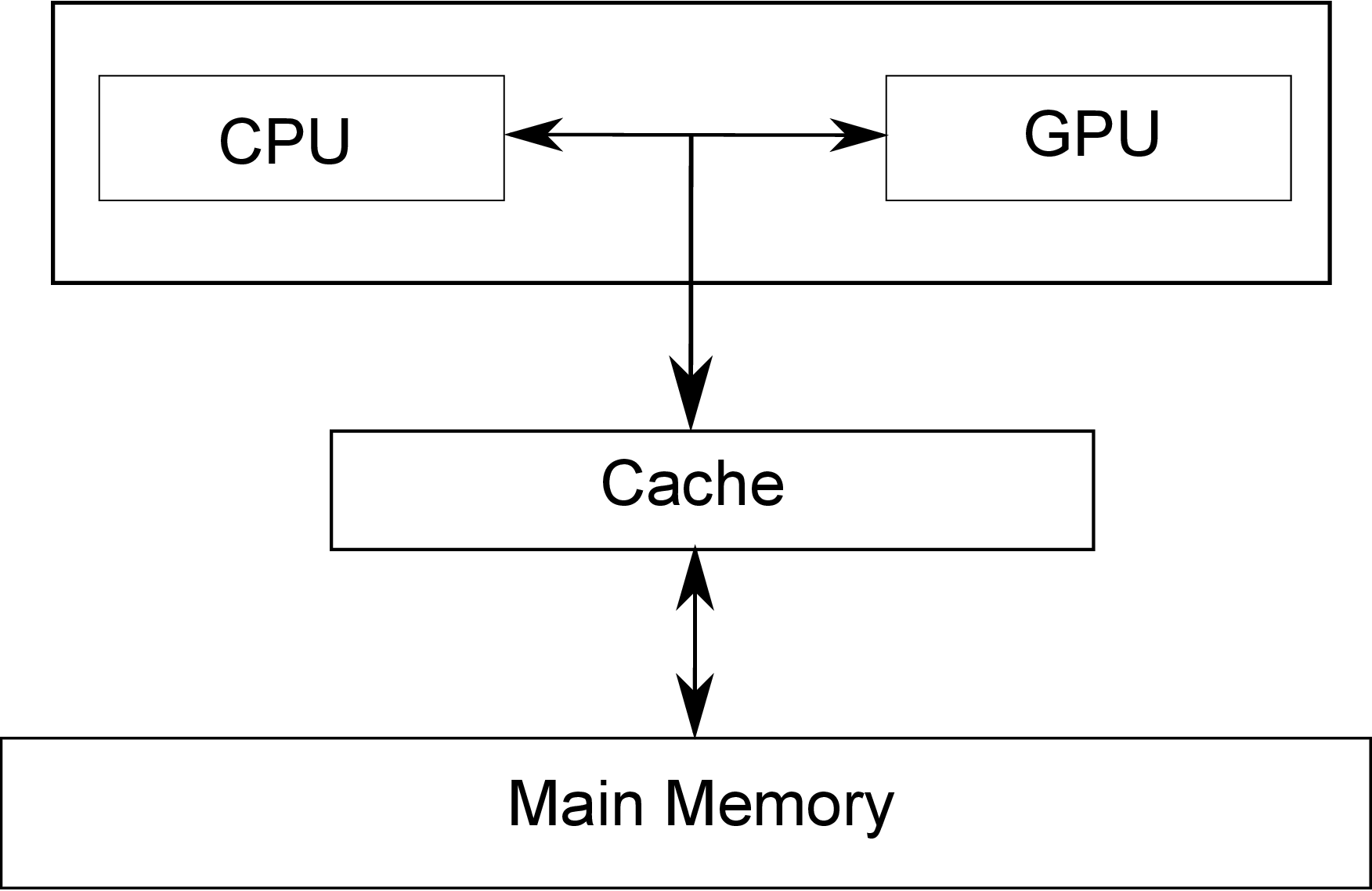}
\label{fig:generalCoupledSystem}}}  \vspace{-2ex}\caption{An
overview of discrete and coupled CPU-GPU architectures}
\label{fig:generalCGsystem}\vspace{-2ex}
\end{figure}

One hurdle for the effectiveness of CPU-GPU co-processing is that
the GPU is usually connected with the CPU with a PCI-e bus, as
illustrated in Figure~\ref{fig:generalCGsystem} (a). On such
discrete architectures, the mismatch between the PCI-e bandwidth
(\hejiong{e.g., 4--8 GB/sec}) and CPU/GPU memory bandwidth
(\hejiong{e.g., dozens to hundreds of GB/sec}) can \hejiong{offset}
the overall performance of CPU-GPU co-processing. As a result, it is
preferred to have coarse-grained co-processing to reduce the data
transfer on the PCI-e bus. Moreover, \hejiong{as} the GPU and the
CPU have their own memory controllers and caches (such as L2 data
cache), data accesses are in different paths.

Recently, vendors have released new coupled CPU-GPU \hejiong{architectures},
such as the AMD APU and Intel Ivy Bridge. An abstract view of the
coupled architecture is illustrated in
Figure~\ref{fig:generalCGsystem} (b). The CPU and the GPU are
integrated into the same chip. They can access the same main memory
space, which is managed by a unified memory controller. Furthermore,
both processors share the L2/L3 data cache, which potentially
increases the cache efficiency.

Table~\ref{table:APUconfiguration} gives an overview of AMD APU
A8-3870K, which is used in our study. We also show the specification
of the latest AMD GPU (Radeon HD 7970) in discrete architectures for
comparison. The GPU in the coupled architecture has a much smaller
number of cores at \hj{lower clock frequency}, mainly because of chip
area limitations. On current AMD APUs, the system memory is further
divided into two parts, which are \emph{host memory} for the CPU and
\emph{device memory} for the GPU. Both of the two memory spaces can
be accessed by either the GPU or the CPU through the \textit{zero
copy buffer}. This study stores the data in the zero copy buffer to
fully take advantage of co-processing capabilities of the coupled
architecture. \saven{The current zero copy buffer is relatively
small, which can be relaxed in the future coupled CPU-GPU
architecture~\cite{AMDdevsummit}.}

\begin{table}
    \centering
\caption{Configuration of AMD Fusion A8-3870K. The last column shows
the configuration of AMD Radeon HD 7970 for reference.}
\label{table:APUconfiguration}
\begin{small}
\begin{tabular}{|c|p{1.1cm}|p{1.1cm}||p{1.4cm}|}
    \hline  & CPU (APU) & GPU (APU)& GPU (Discrete) \\ \hline
    \# Cores & 4 & 400 &  2048 \\\hline
    Core frequency(GHz) & 3.0 & 0.6 & 0.9\\ \hline
    Zero copy buffer (MB) & \multicolumn{2}{c||}{512
(shared)}& --\\\hline
    Local memory size(KB) & 32 & 32 & 32\\ \hline
    Cache size(MB) & \multicolumn{2}{c||}{4
(shared)} & --\\\hline
\end{tabular}
    \centering
    \end{small}
\end{table}

There have been some studies (like MapReduce~\cite{apumapreduce} and
key-value stores~\cite{keyvalue}) on the coupled architecture. Most
studies have demonstrated the performance advantage of the coupled
architecture over the CPU-only or the GPU-only algorithm. This study
focuses on hash joins, and goes beyond existing
studies~\cite{apumapreduce,keyvalue} in two major aspects. \hejiong{Firstly}, we
revisit the design space of hash joins on the coupled architecture
and develop a cost model that can guide the decisions for
co-processing. We conjecture that the design space and cost models
are also applicable to those studies~\cite{apumapreduce,keyvalue}.
\hejiong{Secondly}, we quantitatively show the advantage of co-processing on the
coupled architecture, in comparison with that on the discrete
architecture.

\subsection{OpenCL}

We develop our co-processing schemes based on OpenCL, which is
specifically designed for heterogeneous computing. The advantage of
OpenCL is that the same OpenCL code can run on both the CPU and the
GPU without modification. \hejiong{A} vendor-specific compiler is
employed to optimize OpenCL to the target architecture.
\saven{Previous studies (such
as~\cite{cudaopenclcomparison,openmpopenclcomparison}) have shown
that implementations with OpenCL achieve very close performance to
those with native languages such as CUDA and OpenMP on the GPU and
the CPU, respectively.}

In OpenCL, the CPU and the GPU can be viewed to have the same
logical architecture, namely \emph{compute device}. \saven{On the
APU, the CPU and the GPU are programmed as two compute devices.} A
compute device consists of a number of Compute Units (CUs).
Furthermore, each CU contains multiple Processing Elements (PEs)
running in SIMD style. The piece of code executed by a specific
device is called a \textit{kernel}. A kernel employs multiple
\textit{work groups} for the execution, and each work group contains
a number of \textit{work items}. A work group is mapped to a CU, and
multiple work items are executed concurrently on the CU. The
execution of a work group on the target architecture is vendor
specific. AMD usually executes 64 work items in a \emph{wavefront}
and NVIDIA with 32 work items in a \emph{warp}. All work items in
the same wavefront run in the SIMD manner. In this paper, we use the
terminology from AMD.


OpenCL exposes a \emph{logical} memory hierarchy \hejiong{with} three levels,
i.e., \textit{global memory}, \textit{local memory} and
\textit{private memory}. The global memory is accessible to all work
items with high access latency. The small but fast local memory is
shared by all work items within the same work group. Furthermore,
the smallest private memory is only accessible to a work item with
the lowest latency. The logical memory hierarchy is mapped to the
physical memory hierarchy of the target processor by the compiler.

\subsection{Architecture-aware Hash Joins}

Hash joins are considered as the most efficient join algorithm for
main memory databases. Fruitful research efforts have been devoted
to hash joins in the past decades. Researchers have optimized hash
join algorithms in two \hejiong{main }categories: for a single hash
join~\cite{CacheConsciousHashJoin, hetodscacheoblivious,
databaseNewBottleneck, hashJoinPrefetching, cstore,
ross2007,balkesen13} and for a pipeline of hash joins
\cite{Lo:1993:OPA:170035.170053,Chen:1992:USR:645918.672489}. In
this study, we focus on a single hash join, which can be the basic
building block for a pipeline of hash joins. There are also many
studies on parallel hash joins on multi-processor environments(e.g.,
~\cite{Schneider:1989:PEF:67544.66937,DeWitt:1985:MHJ:1286760.1286774}).
We focus on the related work on modern architectures.

Memory stalls have been a major performance bottleneck for hash
joins in main memory databases~\cite{Ailamaki1999, Manegold2000},
because of random memory accesses. New algorithms (either
cache-conscious~\cite{CacheConsciousHashJoin, databaseNewBottleneck}
or cache-oblivious~\cite{hetodscacheoblivious}) have been developed
\hejiong{to exploit data access locality}.
The other approach~\cite{hashJoinPrefetching} is to hide the memory latency with computation
\hejiong{by software prefetching techniques}.


For multi-core CPUs, memory optimizations continue to be a major
research focus. Garcia et
al.~\cite{pipelinedHashJoinOnMultithreadedArchi} examined a
pipelined hash join implementation. NUMA memory systems have also
been investigated~\cite{Teubner:2011:SPS:1989323.1989389,YinanCIDR}.
In addition to memory optimizations, there have been studies with
architecture-aware tuning and optimizations. Blanas et
al.~\cite{hashJoinOnMultiCoreCPU} showed that synchronization is
also an important factor affecting the overall performance of hash
joins on multi-core processors. Ross~\cite{ross2007} proposed a hash
join implementation based on Cuckoo hashing employing branch
instruction elimination and SIMD instructions. Balkesen et
al.~\cite{balkesen13} carefully evaluated hardware-conscious
techniques.


In addition to CPUs, new architectures (e.g., network
processors~\cite{Gold:2005:ADO:1114252.1114260},
cell~\cite{ross2007} and GPUs~\cite{databaseOperatorsOnGPU,
sortHashRevisited}) have also been proposed to improve join
performance. Query co-processing on GPUs has received a lot of
attentions. \hj{By exploiting} the hardware feature of GPUs, the performance
of hash joins can be significantly
improved~\cite{databaseOperatorsOnGPU,sortHashRevisited}. There have
also been proposals on reducing the PCI-e bus overhead for join
co-processing on the GPU. Kaldewey et al.~\cite{GPUJoinRevisited}
evaluated the join processing on NVIDIA GPUs by adopting UVA
(Universal Virtual Addressing). Pirk et
al.~\cite{hashjoinOnCGwithPCIeconstraint} proposed to exploit the
asymmetric memory channels.

Among various hash join algorithms, there are two basic forms:
simple hash join and partitioned hash join. These two algorithms
have demonstrated very competitive performance on multi-core CPUs
and GPUs in many previous studies~\cite{hashJoinOnMultiCoreCPU,
sortHashRevisited, databaseOperatorsOnGPU}. Thus, this study will
experimentally revisit both of them on the coupled architecture. 

%
%

\section{Hash Join Co-Processing}\label{sec:design}

In the Introduction, we have described the two implications on the
effectiveness of co-processing on the coupled architecture. On the
coupled architecture, co-processing should be fine-grained, and
schedule the workloads carefully to the CPU and the GPU. Moreover,
we need to consider the memory specific optimizations for the shared
cache architecture and memory systems exposed by OpenCL.

We start \hejiong{by} defining the fine-grained \emph{step} definitions in
hash joins. A step consists of computation or memory accesses on a
set of input or intermediate tuples. Next, we revisit the design
space of co-processing, leading to a number of hash join variants
with and without partitioning. Finally, we present the
implementation details of some design tradeoffs in memory
optimizations and OpenCL-related optimizations.

%
%
%

\subsection{Fine-grained Steps in Hash Joins}\label{subsec:finegrained}

A hash join operator works on two input relations, $R$ and $S$. We
assume that $|R| < |S|$. A typical hash join algorithm has three
phases: partition, build, and probe. The partition phase is
optional, and the simple hash join does not have \hejiong{a} partition phase. We
define the fine-grained steps for the simple hash join (SHJ) and the
\hj{partitioned} hash join (PHJ) in Algorithms~\ref{alg:SHJ}
and~\ref{alg:PHJ}, respectively. The granularity of our step
definition is similar to that in the previous study by Chen et
al.~\cite{hashJoinPrefetching}. We will discuss other step
definitions in Section~\ref{subsec:details}.

In SHJ, the build phase constructs an in-memory hash table for $R$.
Then in the probe phase, for each tuple in $S$, it looks up the hash
table for matching \hejiong{entries}. Both the build and the probe phases are divided
into four steps, $b_1$ to $b_4$ and $p_1$ to $p_4$, respectively.
The simple hash join has poor data locality if the hash table cannot
fit into the cache. However, a recent study by Blanas et al.
\cite{hashJoinOnMultiCoreCPU} showed that SHJ is very competitive to
other complex hash join algorithms on the multi-core CPUs,
especially with data skew.

We adopt the hash table \hj{implementation} that has been used in the previous
studies~\cite{hashJoinOnMultiCoreCPU, databaseOperatorsOnGPU,
sortHashRevisited}. A hash table consists of an array of bucket
headers. Each bucket header contains two fields: total number of
tuples within that bucket and the pointer to \hj{a} key list. The key
list contains all \hejiong{the unique} keys with the same hash value, each of
which links a $\mathit{rid}$ list storing the IDs for all tuples
with the same key.


\begin{algorithm}
\caption{Fine-grained steps in SHJ} \label{alg:SHJ}
\begin{small}
\begin{algorithmic}
    \STATE {\small /*build*/}
    \FOR{each tuple in $R$}
    \STATE ($b_1$) compute hash bucket number;
    \STATE ($b_2$) visit the hash bucket header;
    \STATE ($b_3$) visit the hash key lists and create a key header if necessary;
    \STATE ($b_4$) insert the record id into the $\mathit{rid}$ list;
    \ENDFOR
    \STATE {\small /*probe*/}
    \FOR{each tuple in $S$}
    \STATE ($p_1$) compute hash bucket number;
    \STATE ($p_2$) visit the hash bucket header;
    \STATE ($p_3$) visit the hash key lists;
    \STATE ($p_4$) visit the matching build tuple to
compare keys and produce output tuple;
    \ENDFOR
\end{algorithmic}
\end{small}
\end{algorithm}

\begin{algorithm}
\caption{Fine-grained steps in PHJ} \label{alg:PHJ} {\bf Main
Procedure for PHJ: }
\begin{small}
\begin{algorithmic}
    \STATE {\small /*Partitioning: perform multiple passes if necessary*/}
    \STATE \emph{Partition} ($R$);
    \STATE \emph{Partition} ($S$);
    \STATE {\small /*Apply SHJ on each partition pair*/}
    \FOR{each partition pair $R_i$ and $S_i$}
    \STATE Apply SHJ on $R_i$ and $S_i$;
    \ENDFOR
\end{algorithmic}
{\bf Procedure: } \emph{Partition} ($R$)
\begin{algorithmic}
    \FOR{each tuple in $R$}
    \STATE ($n_1$) compute partition number;
    \STATE ($n_2$) visit the partition header;
    \STATE ($n_3$) insert the $<$key, rid$>$ into partition;
    \ENDFOR
\end{algorithmic}
\end{small}
\end{algorithm}

We adopt radix hash join~\cite{databaseNewBottleneck} as PHJ in this
study. In PHJ, it first splits the input relations $R$ and $S$ into
the same number of partitions through a radix partitioning
algorithm. The radix partitioning is performed by multiple passes
based on a number of lower bits of the integer hash values. The
number of passes is tuned according to \hejiong{the} memory
hierarchy (e.g., TLB and data caches). On each pass of partitioning,
the steps are the same, $n_1$ to $n_3$. We use a structure similar
to the hash table to store all the partitions, where a bucket is
used to store a partition. Next, PHJ performs SHJ on each partition
pair from $R$ and $S$ in Algorithm~\ref{alg:SHJ}.

We consider a hash join algorithm with multiple series of
data-parallel steps (namely \emph{step series}) separated by
barriers. For each input data item (e.g., a tuple or
larger-granularity data), it goes through all the steps to generate
the result, and data dependency exists \hejiong{between} two consecutive steps.
We view the step definition with the granularity of one tuple as
\emph{fine-grained} ones. That is, we develop co-processing schemes
based on the per-tuple step definitions in Algorithms~\ref{alg:SHJ}
and~\ref{alg:PHJ}. An SHJ has two step series, $b_1$, ..., $b_4$ for
the build phase and $p_1$, ..., $p_4$ for the probe phase. There is
a barrier between the build and the probe phases. Similarly, \hejiong{a PHJ
with $g$-pass partitioning has $g$ step series ($n_1$, ..., $n_3$)
plus two step series the same as SHJ for joining each partition
pair}. 




\subsection{Revisiting Co-processing Mechanisms}
We revisit the following co-processing mechanisms, which have their
roots in query processing techniques in previous
studies~\cite{databaseOperatorsOnGPU,GPUJoinRevisited,sortHashRevisited}.
Particularly, we describe the high-level idea of each co-processing
mechanism and their strengths and weakness in co-processing on the
coupled architecture. Applying those co-processing schemes to SHJ
and PHJ creates an interesting design space for co-processing on the
coupled architecture. Thus, we summarize a number of hash join
variants at the end of this subsection.

In the following description, we consider a general step series with
$n$ steps denoted \hejiong{by} $s_1$, ..., $s_n$.

{\bf Off-loading (OL).} In the previous study on GPU-accelerated
query processing~\cite{databaseOperatorsOnGPU}, OL is the major
technique to exploit the GPU. For example, the previous
study~\cite{databaseOperatorsOnGPU} proposed to off-load some heavy
operators like joins to the GPU while other operators in the query
\hejiong{remain }on the CPU. The basic idea of OL on a step series
is: the GPU is designed as a powerful massively parallel query
co-processor, and a step is evaluated entirely by either the GPU or
the CPU. Query processing continues on the CPU until the off-loaded
computation completes on the GPU, and vice versa. That is, given a
step series $s_1$, ..., $s_n$, we need to decide if $s_i$ is
performed on the CPU or the GPU.

On the discrete architecture, the PCI-e data transfer overhead is an
important consideration for OL. The decision \hejiong{to off-load a step}
affects the \hejiong{decisions} on other steps. \hejiong{As a result}, we have to
consider $2^n$ possible offloading schemes for the step series. In
contrast, on the coupled architecture, by eliminating the data
transfer overhead, the offloading decision is relatively simple:
depending \hejiong{only }on the performance comparison of running the steps on the
CPU and the GPU.


{\bf Data dividing (DD).} OL could under-utilize the CPU when the
off-loaded \hejiong{computations} are being executed on the GPU, and
vice versa. As the performance gap between the GPU and the CPU on
the coupled architecture \hejiong{is} smaller than that on discrete
\hj{architectures}, we need to keep both the CPU and the GPU busy to
further improve the performance. Thus, the CPU and the GPU work
simultaneously on the same step. We can model the CPU and the GPU as
two independent processors, and the problem is to schedule the
workload to them. This problem has its root in parallel query
processing~\cite{DeWitt:1992:PDS:129888.129894}. One of the most
commonly used schemes is to partition the input data among
processors, perform parallel query processing on individual
processors and merge the partial results from individual processors
as the final result. We adopt this scheme to be \hejiong{the}
data-dividing co-processing scheme (DD) on the coupled architecture.
Particularly, given a step series $s_1$, ..., $s_n$, we need to
decide a \emph{work ratio} for the CPU, $r$ ($0\le r\le 1$) so that
a portion of $r$ of all tuples on the step series are performed on
the CPU, and the remainder are performed on the GPU. If $r=0$, DD
becomes \hejiong{a} CPU-only execution; if $r=1.0$, it becomes
\hejiong{a }GPU-only execution. To reduce the amount of the
intermediate result, each tuple goes through all the steps in a
pipelined manner in DD.


The advantage of DD is that we can leverage the parallel query
processing techniques in parallel databases to keep both the CPU and
the GPU busy. However, DD may still cause under-utilization of the
CPU and the GPU. Considering \hejiong{the fact that }DD uses
coarse-grained workload scheduling, some steps assigned to the CPU
actually have higher performance on the GPU, and vice versa.
\hejiong{That means,} the workload scheduling has to consider the
performance characteristics of the CPU and the GPU, and to perform
fine-grained scheduling for the optimal efficiency on individual
processors.

{\bf Pipelined execution (PL).} To address the limitations of OL and
DD, we consider fine-grained workload scheduling \hejiong{between} the CPU and
the GPU so that \saven{we can capture their performance differences
in processing the same workload. For example, the GPU is much more
efficient than the CPU on $b_{1}$ and $p_{1}$ whereas $b_{3}$ and
$p_{3}$ \hj{are} more efficient on the CPU.} Meanwhile, we should keep
both processors busy. Therefore, we leverage the concept of
pipelined execution and develop an adaptive fine-grained
co-processing scheme for maximizing the efficiency of co-processing
on the coupled architecture.

\begin{figure}
    \centering
    \includegraphics[width=0.32\textwidth]{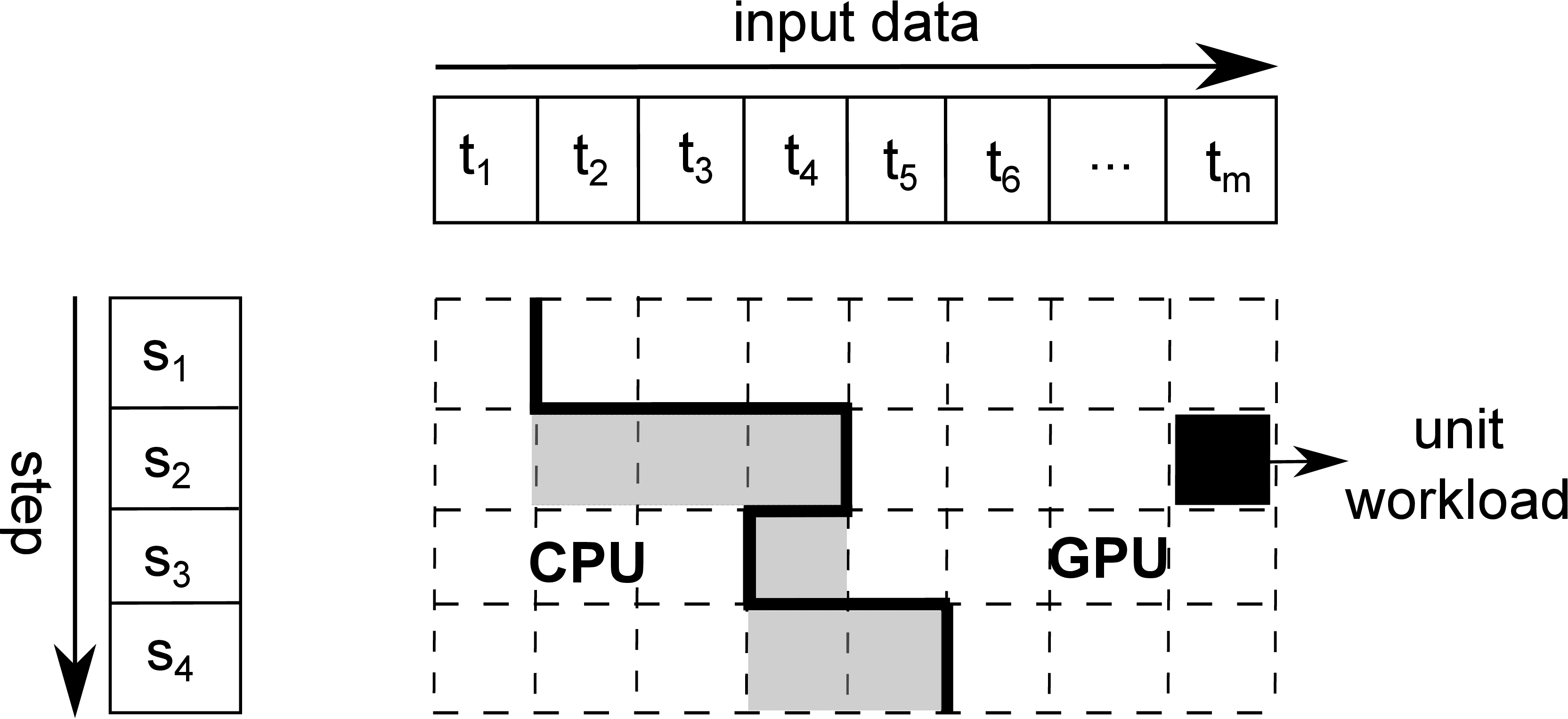}
    \vspace{-2ex}
    \caption{Fine-grained co-processing algorithm on a series of steps. }\label{fig:generalFinegrainedAlgorithm}
    \vspace{-2ex}
 \end{figure}

Unlike DD that has the same workload ratios for all steps, we
consider the data-dividing approach at the granularity of each step
(as illustrated in Figure~\ref{fig:generalFinegrainedAlgorithm}).
Within Step $s_i$, we have a ratio $r_i$ of the tuples assigned to
the CPU and the remainder ($1-r_i$) assigned to the GPU. \hejiong{This differs}
from DD \hejiong{as }we may have different workload ratios across steps.
Intermediate results are generated on two consecutive steps if they
have different workload ratios. For given workload ratios, we prefer
a longer pipeline (a larger number of steps that a tuple can go
through) to reduce the amount of intermediate results. On the other
hand, we need to pay attention to the data dependency \hejiong{between} steps.
If the output data of Step $s_i$ is not available for Step
$s_{i+1}$, Step $s_{i+1}$ has to be stalled.

To achieve the optimal performance, the suitable workload ratios
$r_{i}$ ($1\le i\le n$) should adapt to the computational
characteristics of the steps. We have developed a cost model to
evaluate the performance of the pipelined execution for the given
ratios $r_1$, ..., $r_n$ for Steps $s_1$, ..., $s_n$, respectively.
Then, we use a simple approach to obtain the suitable ratios for the
best estimated performance. Specifically, we consider all the
possible ratios at the step of $\delta$ for $r_i$ ($1\le i\le n$,
$0\le \delta \le 1$), i.e., $r_i$=$\delta$, $2\delta$, $3\delta$,
..., $\lfloor \frac{1}{\delta}\rfloor \delta$. For each set of given
ratios, we use the cost model to predict the performance, and thus
get the suitable workload ratios. In our experiments, we use
$\delta=0.02$ as a tradeoff between the
effectiveness and the execution time of optimizations.


We can consider OL and DD \hejiong{as} special cases for PL. DD is equivalent
to PL with all the ratios the same on all steps. OL is equivalent to
PL with all the workload ratios being either zero or one.

The co-processing schemes presented above outline an interesting
design space for hash joins on the coupled architecture. In this
experimental study, we focus on evaluating how those co-processing
mechanisms are implemented in SHJ and PHJ.

We apply DD, OL and PL schemes to SHJ and obtain three variants
(SHJ-DD, SHJ-OL and SHJ-PL, respectively). For each variant, we
consider all $b_i$ ($1\le i\le 4$) in the build phase and all $p_i$
($1\le i\le 4$) in the probe phase as two different step series.
There are some tuning parameters, and we use the cost model
presented in the next section to determine their suitable values for
the best performance.

Some details of those SHJ variants are described as follows: (1)
\emph{SHJ-DD}: It decides two workload ratios, $r_b$ and $r_p$,
$0\le r_b, r_p\le 1$, for the build and the probe phases,
respectively. For example, in the build phase, a portion of $r_b$ of
the build table are processed by the CPU to construct the hash
table. (2) \emph{SHJ-OL}: It decides whether each of the steps
($b_i$ in the build phase and $p_i$ in the probe phase) is performed
on the CPU or the GPU. (3) \emph{SHJ-PL}: It determines the suitable
workload ratios for all the steps in the build and the probe phases.

%
%
%

We apply DD, OL and PL schemes to PHJ and obtain three variants
(PHJ-DD, PHJ-OL and PHJ-PL, respectively). The idea is the same as
the above SHJ variants, except that we consider the steps in each
pass of partitioning as a step series.


\subsection{Implementations and Design Tradeoffs}\label{subsec:details}


There are a few design tradeoffs in memory optimizations and
OpenCL-related optimizations, which often have significant impact on
the performance of hash joins.


{\bf Memory allocator.} In the hash join algorithm, dynamic memory
\hj{allocations are} common, including (1) the output buffer for a
partition, (2) allocating a new key in the key list, and (3) the
join result output. Current OpenCL (version 1.2) does not support
dynamic memory \hj{allocation} inside the kernel. The efficiency in
supporting dynamic memory \hj{allocations} is essential for those
operations. Inspired by the previous
study~\cite{Hong:2010:MWP:1854273.1854303}, we develop a software
dynamic memory allocator on a pre-allocated array, and memory
requests are dynamically allocated from that array.


A basic implementation of the memory allocator is to use a single
pointer marking the starting address of the free memory in an
pre-allocated array. The pointer initially points to the starting
address of the array. After serving a memory request, the pointer is
moved accordingly. \saven{We use the atomic operation (specifically
\emph{atomic\_add} in our study) to implement a latch for
synchronization among multiple requests. Additionally, the latch is
applicable at both the local memory and the global memory
operations. } However, this basic implementation could suffer from
the contention of atomic operations, especially for supporting
massive thread parallelism on the GPU.

To reduce the contention, we develop an optimized memory allocator.
The memory allocation is at the granularity of a block. The block
size is a tuning parameter and we will experimentally evaluate its
impact in the experiments. The memory allocator maintains a global
pointer. The memory request is always made by work item 0 (Thread 0)
from a work group, and the global pointer is advanced by one block
size. The memory allocator returns one block to the work group. The
threads within the work group request memory from the block. It uses
a local pointer for synchronization among the threads within the
work group. We allocate the local pointer in the local memory to
reduce the memory access overhead. Note, the local memory is
accessible to all the threads within a work group.

The optimized memory allocator resolves the contention on the GPU.
Additionally, the pre-allocation is also beneficial for the CPU. It
eliminates many small \emph{malloc} operations, and reduces the
memory allocation overhead on the CPU.

{\bf Shared vs. separate hash tables.} Cache performance and
concurrency are two key factors for the design of hash \hj{tables}. An
important design choice in the build phase is to use a shared hash
table or separate hash tables for the CPU and the GPU. Either
solution has its own \hejiong{benefits and drawbacks}. A shared hash table has better
cache performance, since the CPU and the GPU share the data cache on
the coupled architecture. Additionally, it uses less memory space in
the zero copy buffer. On the other hand, separate hash tables have
smaller latch contention \hejiong{between} the CPU and the GPU. We
experimentally evaluate these two solutions.


{\bf Workload divergence.} In OpenCL, all work items in the same
wavefront run simultaneously in a lockstep. The execution time of a
wavefront is equal to the worst execution time of all work items.
Workload divergence among work items causes severe penalties in the
overall performance of the wavefront. One common source \hejiong{of} workload
divergence in hash joins is data skew. For example, data skews cause
workload divergence in $b_3$ in the build phase and $p_3$ in the
probe phase.

To reduce the workload divergence, we adopt a grouping-based
approach in the previous study~\cite{He:2011:HTE:1952376.1952381},
which is used to reduce the branch divergence of transaction
executions on the GPU. In order to reduce the workload divergence,
we group the input data according to the amount of workload. \hejiong{Using}
the probe as an example, the input data are the hash bucket headers
given in $b_2$, and the amount of workload is represented by the
number of keys in the key list. After grouping, the input data with
the similar amount of workload are grouped together, and the
divergence among the work items in a work group is reduced. The
number of groups is tuned for the tradeoff between the grouping
overhead and the gain of reduced workload divergence.

{\bf Step definitions.} So far, we have adopted very fine-grained
step definitions for co-processing. Ideally, there are other
\hj{granularities} of defining steps, which can have different memory
performance and efficiency of co-processing.

We are particularly interested in evaluating the step definition for
PHJ in the previous study~\cite{hashJoinOnMultiCoreCPU}. After the
partition phase, we have got $P$ partition pairs on $R$ and $S$
($<$$R_i$, $S_i$$>$, $0\le i\le{P-1}$). The further join processing
on $R_i$ and $S_i$ is performed by one thread. Thus, all the join
processing on the $P$ partition pairs on $R$ and $S$ can be viewed
as a step, and a partition pair is considered to be the input data
item to the step. Thus, the granularity for a step for this
co-processing is a SHJ on the partition pair. Note, those SHJs use
separate hash tables, which potentially loses the opportunities of
cache reuse in our fine-grained PHJ variants.


\section{Performance Model}\label{sec:model}

There are a number of parameters (for example, the ratio in the
data-dividing scheme) to be determined for the co-processing
performance on the coupled architecture. In this section, we develop
a cost model to predict the performance of hash joins with
co-processing on the coupled architecture, and then use the cost
model to determine the suitable values for the parameters.



Cost models for hash joins have been developed in previous studies,
e.g., on the CPU~\cite{databaseNewBottleneck,
Manegold:2002:GDC:1287369.1287387} and the
GPU~\cite{RelationalQueryCoProcessing}. Those cost models are highly
specialized for the target architectures and the
architecture-specific query processing algorithms. \saven{For
example, the cost model for GPU-based
co-processing~\cite{RelationalQueryCoProcessing} needs the cost
estimation on the PCI-e data transfer.} In this study, the
implementations on the CPU and the GPU are based on the same OpenCL
programs with different input parameters. The CPU and the GPU are
abstracted as the same compute device model. Moreover, the CPU and
the GPU share the data cache and the main memory without the PCI-e
bus on the coupled architecture. It is desirable to have an unified
model to estimate the performance of co-processing on both the CPU
and the GPU \hejiong{in terms of both computation and memory
latency}.


\begin{table}
\centering
\begin{scriptsize}
\caption{Notations in the cost model}\label{table:modelNotation}
\begin{tabular}{|l|p{6.5cm}|}
\hline
Notation & Description\\
\hline
        $r_i$ & Workload ratios for the CPU at Step $i$ ($1\le i\le \#\mathit{Steps}$) \\ \hline
        $x_i$ & The number of input items (e.g., tuples) at Step $i$ ($1\le i\le \#\mathit{Steps}$) \\ \hline
        $T$ & Total elapsed time of executing the entire step series \\ \hline
        $T_{\mathit{XPU}}$ & The execution time on the processor
        $\mathit{XPU}$ ($\mathit{XPU}=\{\mathit{CPU},
        \mathit{GPU}\}$)\\ \hline
        $T^i_{\mathit{XPU}}$ & The execution time on the processor
        $\mathit{XPU}$ at Step $i$\\ \hline
        $C^{i}_{\mathit{XPU}}$ & Computation time at Step $i$ \\ \hline
        $M^i_{\mathit{XPU}}$ & Total global memory access time at Step $i$ \\ \hline
        $D^{i}_{\mathit{XPU}}$ & The pipelined delay at Step $i$ \\ \hline
        $\#\mathit{I}^{i}_{\mathit{XPU}}$ & The average number of instructions on $\mathit{XPU}$ per tuple in Step $i$ \\ \hline
        $\mathit{IPC}_{\mathit{XPU}}$ & The peak instruction per cycle on $\mathit{XPU}$\\ \hline
\end{tabular}
\centering
\end{scriptsize}
\end{table}


In Section~\ref{sec:design}, we have identified PL as a basic
abstraction for hash join co-processing on the coupled architecture
(DD and OL are special cases for PL). This allows us to abstract the
common design issues on the cost model for different co-processing
schemes. We call this high level abstraction the
\emph{abstract model}. For a specific co-processing algorithm, we
obtain its cost model by instantiating the abstract model with
profiling and algorithm-specific analytic modelling (e.g., memory
optimizations). In the remainder of this section, we first present
the abstract model for PL on the coupled architecture, and next use
SHJs as an example of instantiating the abstract model.

\subsection{The Abstract Model}

We model the elapsed time of executing a step series with PL. The
step series consists of $n$ steps, with $x_i$ input data items on
Step $i$ ($1\le i\le n$). The cost model predicts the elapsed time
of PL with workload ratios $r_i$ in Step $i$. The parameters are
summarized in Table~\ref{table:modelNotation} for reference. Thus,
the total elapsed time $T$ is estimated to be the longer execution
time of the CPU and the GPU.
\begin{equation}\label{Eq:totaltime}
T=max(T_{\mathit{XPU}}), \mathit{XPU}=\{\mathit{CPU}, \mathit{GPU}\}
\end{equation}

On each processor, we estimate the execution time to be the total
execution time of all steps in the step series. The execution time
of Step $i$ ($T^i_{\mathit{XPU}}$) is further estimated in three
parts: computation time $C^{i}_{\mathit{XPU}}$, global memory access
time $M^i_{\mathit{XPU}}$ and the pipelined delay
$D^{i}_{\mathit{XPU}}$.
\begin{equation}\label{Eq:overalltime}
T_{\mathit{XPU}} =
\sum\limits_{i=1}^{n}T^i_{\mathit{XPU}}=\sum\limits_{i=1}^{n}
(C^{i}_{\mathit{XPU}}+M^i_{\mathit{XPU}}+D^{i}_{\mathit{XPU}})
\end{equation}
\noindent We describe the estimation of each component accordingly.

We estimate the computation time to be the total execution time
for all the instructions of the step. We assume an ideal execution
pipeline with the optimal IPC on each processor, and a hash join on
uniform data distributions.
\begin{equation}\label{eq:computationTime}
C^{i}_{\mathit{XPU}} = \frac{\#\mathit{I}^{i}_{\mathit{XPU}}\times
r_i\times x_i}{\mathit{IPC}_{\mathit{XPU}}}
\end{equation}

To estimate the memory stalls, we adopt a traditional calibration
method for the CPU~\cite{Manegold:2002:GDC:1287369.1287387} and for
the GPU~\cite{RelationalQueryCoProcessing}, which can estimate the
memory stall cost per data item in each step. The basic idea is to
measure the memory stall cost, by excluding the caching effects and
the thread parallelism. It is suitable to estimate the memory \hj{stall costs}
for both random and sequential accesses. For details on the
calibrations and models, we refer readers to the original
paper~\cite{RelationalQueryCoProcessing,Manegold:2002:GDC:1287369.1287387}.
For different steps, we have instantiated the model with the
consideration of their access patterns.

For $D^{i}_{\mathit{XPU}}$, this delay is caused by different
workload ratios in the two consecutive steps. Due to data dependency
\hejiong{between} the two processors, one processor has to synchronize with the
other one on Step $i$ if the workload ratios are different on Steps
$i$ and $(i-1)$. The delay occurs when the input data for Step $i$
\hj{have} not been generated from Step $(i-1)$. There are two cases
depending on the workload ratio comparison.

Case 1: if $r_i > r_{i-1}$, we use Eq.~\ref{eq:dcase1}. The CPU may
encounter delay while the GPU prepares \hejiong{the} input data in
Step $(i-1)$ for the CPU in Step $i$. \hejiong{The ratio of the GPU
time that is not pipelined with the CPU in Step $(i-1)$ is
$\frac{1-r_i}{1-r_{i-1}}$. By subtracting this amount of time
($T^{i-1}_{GPU}\times \frac{1-r_i}{1-r_{i-1}}$) from the total GPU
time from Step $1$ to Step $i-1$, we get the time of the GPU from
Step $1$ to the end of the pipelined execution area in Step $i-1$.
The total CPU time from Step $1$ to the end of pipelined execution
area with the GPU in Step $i$ is
$\sum\limits_{j=1}^{i}T^{j}_{\mathit{CPU}}$. Therefore,
$D^{i}_{CPU}$ is the delay time on the CPU to wait for data
generated from the GPU in Step $i-1$, as in Eq.~\ref{eq:dcase1}.
Naturally, if $D^{i}_{CPU} \le 0$, it is set as 0, meaning there
will be no pipelined execution delay.} \hejiong{
\begin{equation}\label{eq:dcase1}
\begin{split}
D^{i}_{\mathit{CPU}} = (\sum\limits_{j=1}^{i-1}T^{j}_{\mathit{GPU}}
- T^{i-1}_{GPU}\times \frac{1-r_i}{1-r_{i-1}}) -
\sum\limits_{j=1}^{i}T^{j}_{\mathit{CPU}}
\end{split}
\end{equation}
} 

Case 2: If $r_i < r_{i-1}$, we use Eq.~\ref{eq:dcase2} to calculate
the pipelined execution delay for the GPU in a similar way to Case
1. \hejiong{
\begin{equation}\label{eq:dcase2}
\begin{split}
D^{i}_{\mathit{GPU}} = &\sum\limits_{j=1}^{i-1}T^{j}_{\mathit{CPU}}
-
(\sum\limits_{j=1}^{i}T^{j}_{\mathit{GPU}}-T^{i}_{\mathit{GPU}}\times\frac{1-r_{i-1}}{1-r_i})
\end{split}
\end{equation}
}

Finally, as specific to the pipelined co-processing, we need to
estimate the cost of intermediate results and the communication
cost between the CPU and the GPU. We estimate the cost for two
consecutive steps. If they have the same workload ratio, we can
ignore the costs on intermediate results and the communication.
Otherwise, the amount of intermediate results can be derived from
the difference of the workload ratio between the two steps. For Step
$i$, the number of intermediate data items is $(r_i - r_{i-1})\times
x_i$, assuming a uniform distribution.


\subsection{Model Instantiation}

We use SHJs as an example to illustrate the model instantiation. We
consider fine-grained co-processing SHJ variants (with the
granularity of tuples).

For computation time, the optimal IPC value can be obtained from the
hardware specification. The number of instructions for per tuple in
each step can be obtained from profiling tools such as AMD CodeXL
and AMD APP Profiler. One \hejiong{issue to handle} is
that the number of instructions per tuple in some steps varies with
the workload. For example, the costs of $b_3$ and $p_3$ depend on
the length of the key list. Therefore, we use the number of
instructions per key search as the unit cost for that step, and
estimate the number of instructions of the step to be the number of
instructions per key search multiplied by the average number of keys
in the key list.

Next, we calibrate the memory unit cost per tuple with the
calibration method~\cite{RelationalQueryCoProcessing}. For some
steps, the memory unit cost is workload-dependent. We adopt the same
approach as profiling the number of instructions discussed above.

Given the above calibrations, we can derive the total cost for
different SHJ variants. Here, we use SHJ-DD as an example, and we
can derive the cost for other variants similarly. We estimate the
performance of SHJ-DD as the total elapsed time of the build phase
and the probe phase. Since the estimation of the build phase is
similar to the probe phase, we estimate the build phase as follows.
We instantiate Eqs.~\ref{Eq:totaltime}--\ref{eq:dcase2}. In
Eq.~\ref{Eq:overalltime}, we set $n=4$, because there are four steps
in the build phase. In Eq.~\ref{eq:computationTime}, the number of
instructions \hejiong{is} substituted with the profiled result \hejiong{from} each step.
We then use the calibrated values for each step to get the memory
costs, and determine the pipelined execution delay in
Eqs.~\ref{eq:dcase1} and~\ref{eq:dcase2}. Thus, we can get the cost
for the build phase in Eq.~\ref{Eq:totaltime}.

\section{Evaluation} \label{sec:evaluation}

In section, we experimentally evaluate the co-processing scheme for
the hash join on the coupled CPU-GPU architecture. Overall, there
are four groups of experiments. \hejiong{Firstly}, we investigate
the overhead of data transfers and other operations of co-processing
for the hash join on a discrete CPU-GPU architecture
(Section~\ref{subsec:discreteeval}). \hejiong{Secondly}, we evaluate
the accuracy of our cost model (Section~\ref{subsec:costmodel}).
\hejiong{Thirdly}, we study the performance impact of the design
tradeoffs in co-processing of hash joins
(Section~\ref{subsec:tradeoff}). Finally, we show the end-to-end
performance comparison (Section~\ref{subsec:cop}).

%

\subsection{Experimental Setup}


\textbf{Hardware configuration.} We conduct our experiments on a PC
equipped with an AMD APU A8-3870K. The hardware specification has
been summarized in Table~\ref{table:APUconfiguration}.

\textbf{Data sets.} We use the same synthetic data sets as the
previous study~\cite{hashJoinOnMultiCoreCPU} to evaluate our
implementations. Both relations $R$ and $S$ consist of two four-byte
integer attributes, namely the record ID and key value. Our default
data set size is 16 million tuples with uniform-distributed key
values for both \hj{relations} $R$ and $S$, unless specified otherwise.
Both $R$ and $S$ can be considered as basic relations (without
compression) in column-oriented databases, or the intermediate
relations by extracting the key and record ID from much larger
relations in row-oriented databases. We experimentally evaluate the
impact of different data sizes. In addition to uniform data sets, we
also use skewed data sets and vary the selectivity to show the
overall performance comparisons. \saven{We created two skewed
datasets with $s\%$ of tuples with one duplicate key values:
\emph{low-skew} with $s=10$ and \emph{high-skew} with $s=25$.}

\textbf{Implementation details}. We develop all hash join variants
using OpenCL 1.2. The OpenCL configuration including work groups and
work items has been tuned to fully utilize the CPU and the GPU
capabilities. In addition to co-processing, we also implement
CPU-only and GPU-only algorithms. Our implementations have adopted
the common practice in the previous
implementations~\cite{databaseOperatorsOnGPU,hashJoinOnMultiCoreCPU,databaseNewBottleneck}.
For example, we choose the hash function MurmurHash 2.0 that is also
used in the previous study~\cite{hashJoinOnMultiCoreCPU}, which has
\hejiong{a }good hash collision rate
and low computational overhead.



We compare the discrete and the coupled architectures to investigate
the performance issues of discrete architectures. For a fair
comparison, we should ideally use the CPU and the GPU that have
exactly the same architectures as those in the coupled architecture.
However, this may be not always feasible as the coupled architecture
advances. Instead, we use the CPU and the GPU on the coupled
architecture to emulate the CPU and the GPU on \hj{a} discrete
architecture, and emulate the PCI-e bus data transfer with a delay.
This simulation-based approach allows us to evaluate the impact of
different processors and PCI-e bus. \hejiong{Unlike} real discrete
architectures, the CPU and the GPU in the \hejiong{emulated} architecture still
share the cache. In our experiment, we find that the \hejiong{emulated}
approach is able to help us to understand the performance issues of
the discrete architecture. The delay of one data transfer on PCI-e
bus is estimated to be
$\mathit{latency}+\frac{\mathit{size}}{\mathit{bandwidth}}$, where
$\mathit{size}$ is the data size and $\mathit{latency}$ and
$\mathit{bandwidth}$ \hj{are} the latency and bandwidth of the PCI-e bus,
respectively. In our study, we emulate a PCI-e bus with
$\mathit{latency}=0.015$ ms and $\mathit{bandwidth}=3$ GB/sec.

On the emulated discrete architecture, we can implement all the hash
join variants except PL. \hejiong{Using} SHJ-DD as an example\hejiong{, in} the build
phase, a part of the build table is transferred to the GPU memory
before the GPU starts building hash tables. An estimated delay is
added so that the GPU starts the build phase later. When the build
phase is done, the partial hash table is transferred back to the CPU
for a merge operation. The probe phase has a similar process to the
build phase. For PL, it is inefficient to implement the fine-grained
co-processing on the discrete architecture for two reasons. \hejiong{Firstly},
there are much more memory data transfers by exchanging the
intermediate results between the CPU and the GPU. \hejiong{Secondly}, the data
dependency control logic on the GPU memory is difficult to implement
on the PCI-e bus. Thus, we use the execution of PL on the coupled
architecture to analyse the impact of PL on the discrete
architecture, including intermediate results and efficiency of
pipelined execution.

%

\subsection{Evaluations on Discrete
Architectures}\label{subsec:discreteeval}


\begin{figure}[!htb]
    \centering
    \includegraphics[width=0.35\textwidth]{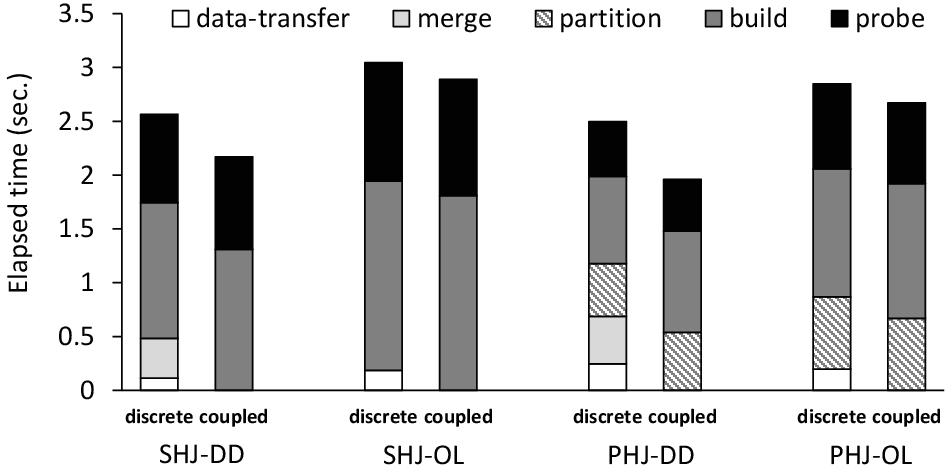}
    \vspace{-2ex}
    \caption{Time breakdown on discrete and coupled
    architectures
    }\label{fig:timebreakdownofstage}
    \vspace{-2ex}
\end{figure}

\textbf{Data transfer and merge overhead.} We first study the time
breakdown of hash join variants on the discrete architecture. For
comparison, we also show the time breakdown on the coupled
architecture. Figure~\ref{fig:timebreakdownofstage} shows the time
breakdown for the data dividing and off-loading co-processing
schemes. On the discrete architecture, the workload ratios for the
build and probe phases for SHJ-DD are 25\% and 42\%, respectively;
the workload ratios for the partition, build and probe phases for
PHJ-DD are 11\%, 26\% and 41\%, respectively. We can see that the
suitable workload ratios vary \hj{across} different phases. SHJ-OL and
PHJ-OL have degraded to be GPU-only, because all the steps on the
GPU are faster than those on the CPU (as we will discussed later).

On the discrete architecture, co-processing usually requires
explicit data transfer on the PCI-e bus and the merge operation. In
the experiments, DD has both kinds of overhead and OL has only the
data transfer overhead because OL is essentially GPU-only. Overall,
the PCI-e data transfer overhead is 4--10\% of the overall execution
time. PHJ generally has higher data transfer overhead, since it has
one more phase than SHJ. In comparison, this data transfer overhead
is eliminated on the coupled CPU-GPU architecture.

The merging operation usually creates a more significant overhead
than the data transfer on PCI-e bus (14\% and 18\% of the overall
time in SHJ-DD and PHJ-DD, respectively). While the data transfer
overhead on \hejiong{the }PCI-e bus can be reduced with advanced overlapping \hejiong{of}
computation and data transfer, the merge overhead is inherent to DD
on the discrete architecture. In comparison, on the coupled
architecture, DD uses a shared hash table (as we evaluate in
Section~\ref{subsec:tradeoff}), and the merge overhead is
eliminated.




\textbf{Workload ratios in DD.} We study the workload ratios on two
different architectures (Figures are omitted due to space limits).
Generally, we find that the suitable workload ratios \hejiong{of} the CPU on
the discrete architectures are higher than those on the coupled
architecture. During the data transfer, the CPU can do more useful
work for DD. Since the data transfer accounts for 4--10\% of the
total execution time, the difference is small (smaller than 5\%).


\begin{figure}
    \centering
    \includegraphics[width=0.35\textwidth]{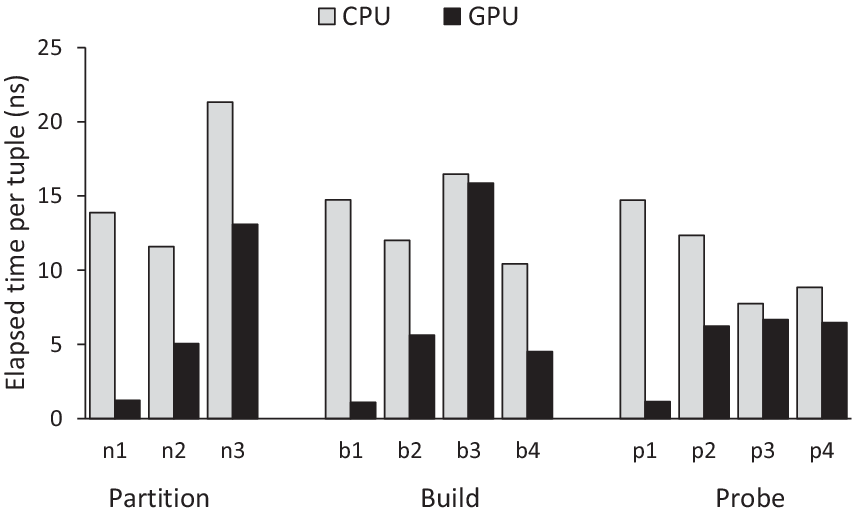}
    \vspace{-2ex}
    \caption{Unit costs for different steps on the CPU and the GPU for PHJ}\label{fig:speedofCGineachstepPHJ}
    \vspace{-2ex}
\end{figure}

\textbf{GPU and CPU processing capabilities for different steps.} In
order to achieve the efficiency of employing fine-grained
co-processing, we need to understand the processing performance of
each step on the CPU and the GPU. We measure the performance of each
step on the CPU-only and the GPU-only algorithms.
Figure~\ref{fig:speedofCGineachstepPHJ} shows the average processing
time per tuple for different steps in PHJ on the CPU and the GPU. We
have observed similar results on SHJ. Overall, the hash value
computation can greatly benefit from the GPU acceleration due to its
massive data parallelism and computation intensity. Therefore, the
GPU can accelerate the hash value computation based steps (i.e.,
$n_1$, $b_1$, and $p_1$) by more than 15X. Instead, some other
operations, such as $b_3$ or $p_3$, cannot match the GPU
architecture features, because of random memory accesses and
divergent branches. As a result, the GPU and the CPU have very close
performance on those steps. Since different steps may have different
performance comparison on the CPU and the GPU, this confirms that a
fine-grained co-processing algorithm is essential to better exploit
the strength of the CPU and the GPU. Workloads of different steps
should be carefully assigned to suitable compute devices. That leads
to our further analysis of PL.

\begin{figure}
\centerline{ \subfigure[build in SHJ
]{\includegraphics[width=0.17\textwidth]{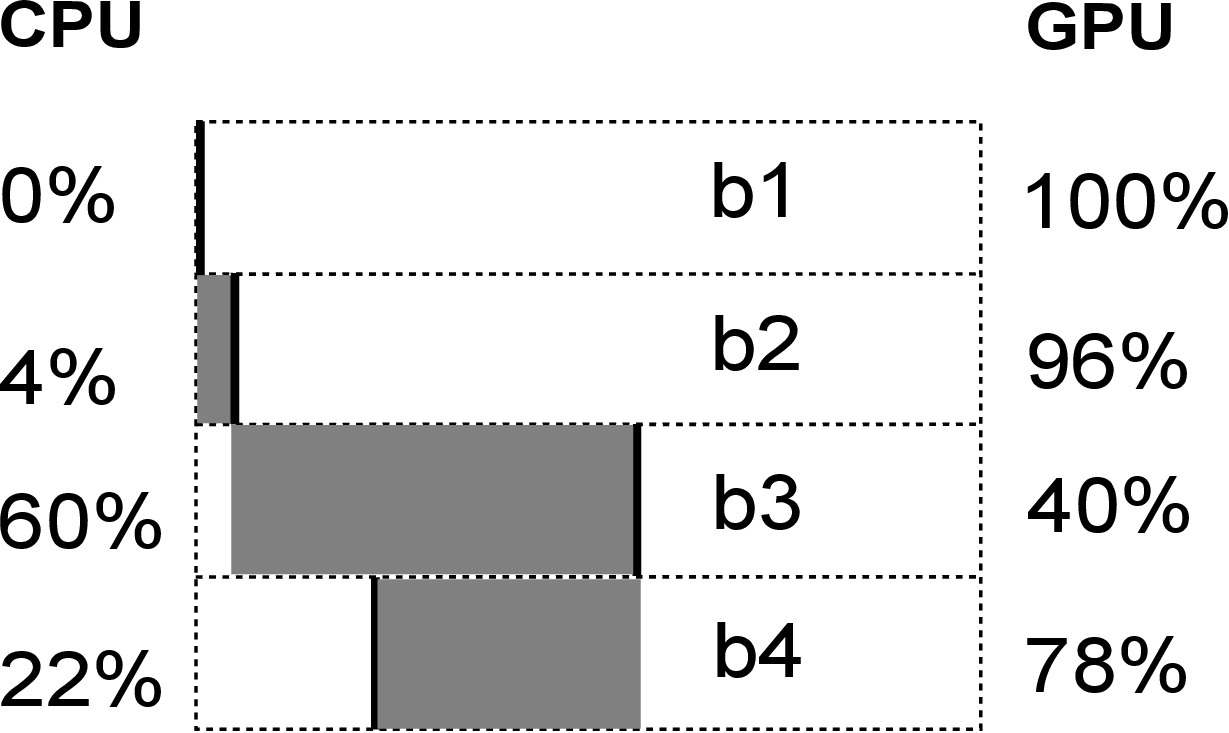}
\label{fig:SHJbuildratio}} \hfil \subfigure[probe in SHJ
]{\includegraphics[width=0.17\textwidth]{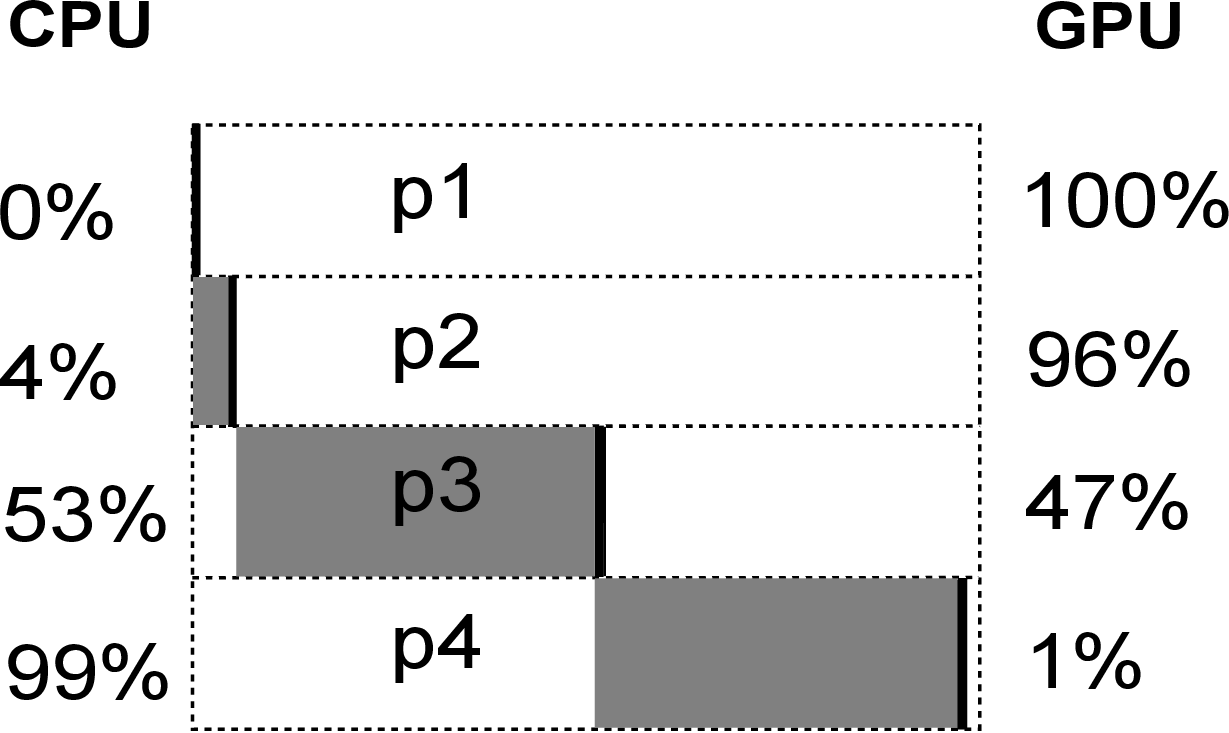}
\label{fig:SHJproberatio}}}

\vspace{-2ex} \caption{Optimal workload ratios of different steps
for SHJ-PL on the coupled architecture} \label{fig:SHJratio}
\vspace{-2ex}
\end{figure}

\begin{figure}
\centerline{ \subfigure[\small{partition in PHJ}]
{\includegraphics[width=0.15\textwidth]{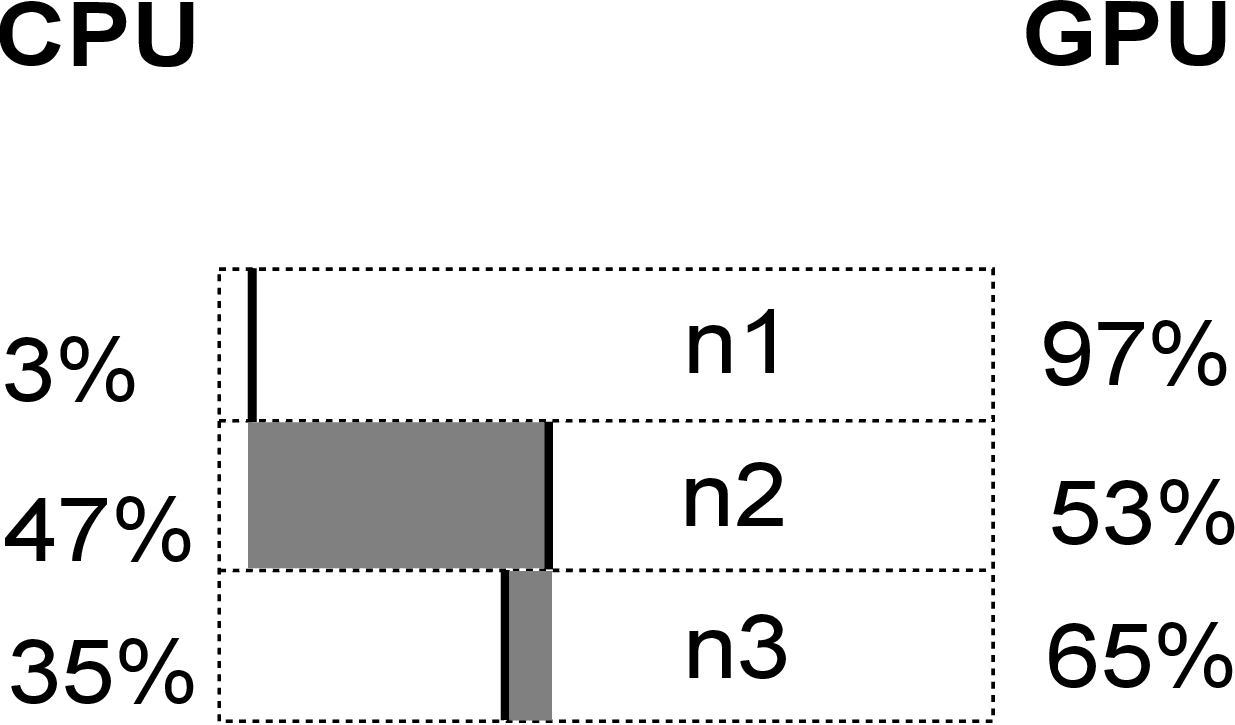}
\label{fig:PHJpartitionratio}} \hfil \subfigure[\small{build in PHJ}]
{\includegraphics[width=0.15\textwidth]{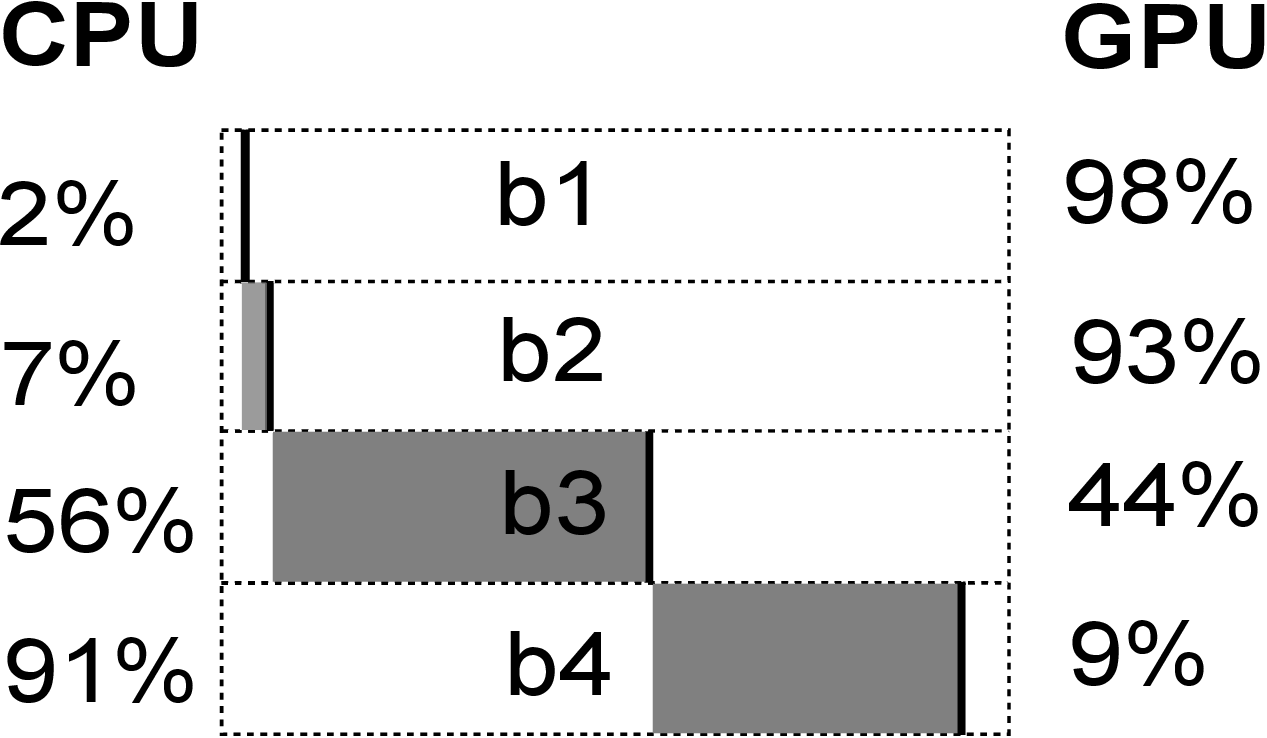}
\label{fig:PHJbuildratio}}     \hfil \subfigure[\small{probe in PHJ}]
{\includegraphics[width=0.15\textwidth]{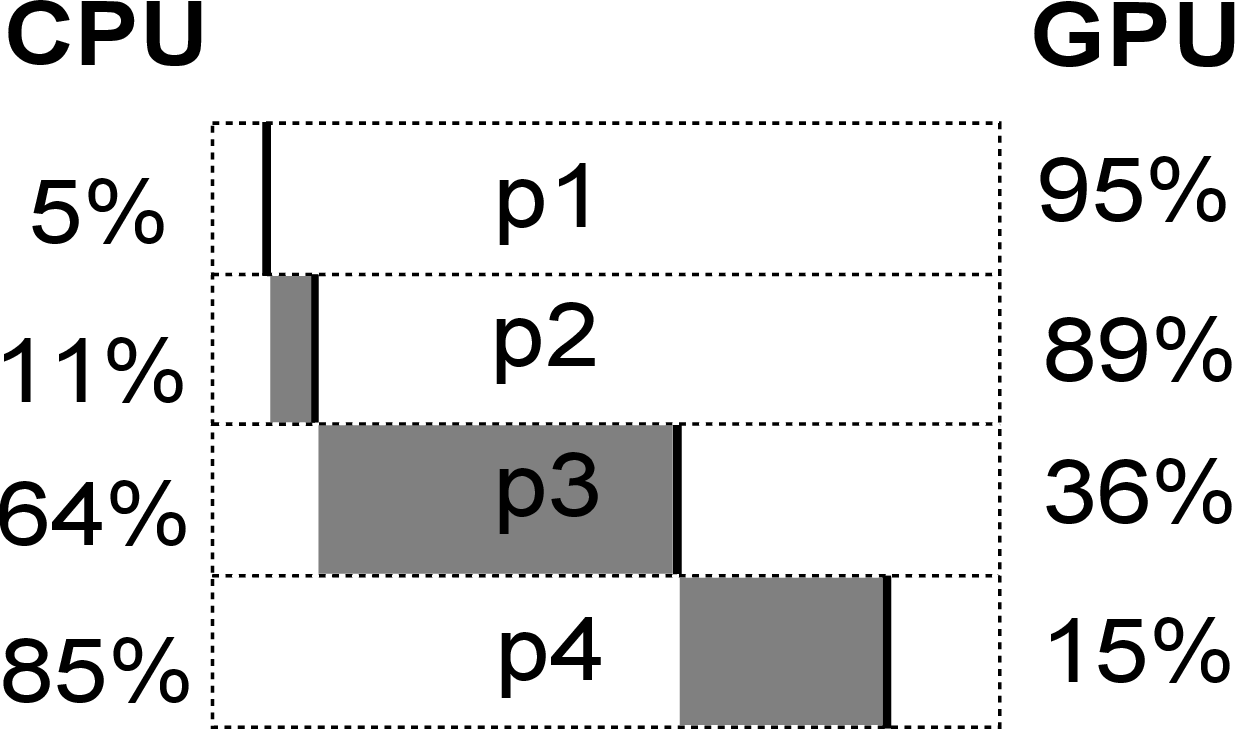}
\label{fig:PHJproberatio}}}  \vspace{-2ex}\caption{Optimal workload ratios of different steps for PHJ-PL on the coupled architecture}
\label{fig:PHJratio}\vspace{-2ex}
\end{figure}

\textbf{Analyzing PL. } We analyze the workload ratios of different
steps for PL on the coupled architecture to understand the impact of
intermediate results and efficiency of pipelined execution. Figures
\ref{fig:SHJratio} and \ref{fig:PHJratio} show the optimal workload
ratios of different steps in SHJ-PL and PHJ-PL on the coupled
architecture, respectively. Overall, the optimal workload ratios are
varied across different steps. The GPU can take all workload ($b_1$
in Figure \ref{fig:SHJbuildratio} and $p_1$ in Figure
\ref{fig:SHJproberatio}), or takes only a very small portion of
workload (1\% for $p_4$ in Figure \ref{fig:SHJproberatio}). We have
the following two implications.

\hejiong{Firstly}, PL can be inefficient on the discrete architecture, since the
significant difference in the workload ratios among the steps
creates significant overhead in data transfer and pipeline
execution. Recall that, in PL, the workload ratio difference \hejiong{between} two
consecutive steps determines the amount of intermediate results. We
use \hejiong{the} grey area in the figures to indicate the intermediate results
to be generated during the execution of PL. For example, in
Figure~\ref{fig:SHJratio}, the CPU only takes 4\% workload for
$b_2$, but takes 60\% for $b_3$. That requires \hejiong{a transfer of} 56\% of
the output data of $b_2$ on the PCI-e bus, if PL runs on the
discrete architecture. On the other hand, the overhead of pipelined
execution is not shown in the figure. The difference in the workload
ratios results in different lengths of pipelines in the execution,
which results in workload divergence.

The other implication is that, since the optimal workload ratios
vary for different steps, the conventional approach (by offloading
the entire join to the GPU) cannot fully exploit the co-processing
capabilities from both the CPU and the GPU. 


\subsection{Cost Model Evaluation}\label{subsec:costmodel}


Overall, our cost model can accurately predict the performance of
all hash join variants with different tuning parameters, and thus
can guide the decision \hejiong{of} getting the suitable value. Since the
results are similar for both SHJs and PHJs, we focus \hejiong{of} the results
on SHJs.

\begin{figure}
\centerline{ \subfigure[build
]{\includegraphics[width=0.23\textwidth]{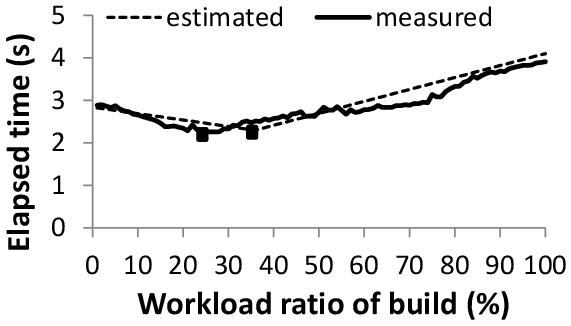}
\label{fig:costmodelDDbuild}} \hfil \subfigure[probe
]{\includegraphics[width=0.23\textwidth]{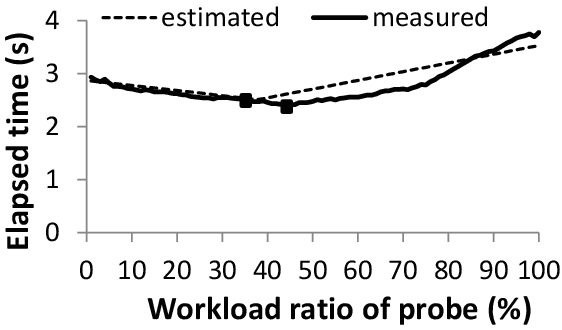}
\label{fig:costmodelDDprobe}}}  \vspace{-2ex}\caption{Estimated and
measured time for SHJ-DD with workload ratios varied}
\label{fig:costmodelDD}\vspace{-2ex}
\end{figure}

\begin{figure}
\centerline{ \subfigure[build
]{\includegraphics[width=0.23\textwidth]{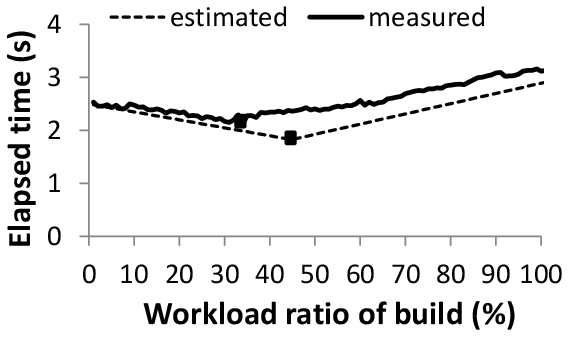}
\label{fig:costmodelOHbuild}} \hfil \subfigure[probe
]{\includegraphics[width=0.23\textwidth]{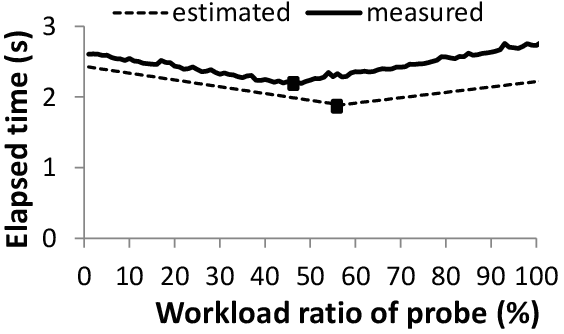}
\label{fig:costmodelOHprobe}}}  \vspace{-2ex}\caption{Estimated and
measured time for a special case for PL: offloading entire $b_1$ and
$p_1$ to the GPU, and applying data-dividing with the same ratio $r$
to all the other steps. The results are measured by varying $r$.}
\label{fig:costmodelOH}\vspace{-2ex}
\end{figure}

Figure~\ref{fig:costmodelDD} compares the estimated performance with
measured performance for SHJ-DD with the workload ratios varied. The black solid squares
in the figures denote the optimal points. It shows that our
estimated time is close to the measured time. However, the estimated time is
slightly lower than \hejiong{the} measured time. One factor \hejiong{to consider }is that our cost model
does not include the estimation of the lock contention overhead.

PL has a large design space. In general, our estimation can closely
predict the performance of PL with different workload ratios. For
clarity of presentation, we consider a special case for PL:
offloading \hejiong{the} entire $b_1$ and $p_1$ to the GPU, and applying
data-dividing with the same ratio $r$ to all the other steps. The
results in Figure~\ref{fig:costmodelOH} are measured by varying $r$.
In this special case of PL, we see that our prediction is close to
the performance varying $r$ and also is able to predict its suitable
value.

\begin{figure}
\centerline{ \subfigure[build of SHJ-PL
]{\includegraphics[width=0.23\textwidth]{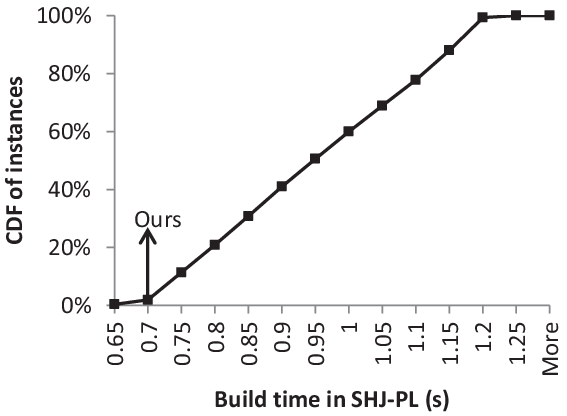}
\label{fig:cdfbuildtimeinshj}} \hfil \subfigure[probe of PHJ-PL
]{\includegraphics[width=0.23\textwidth]{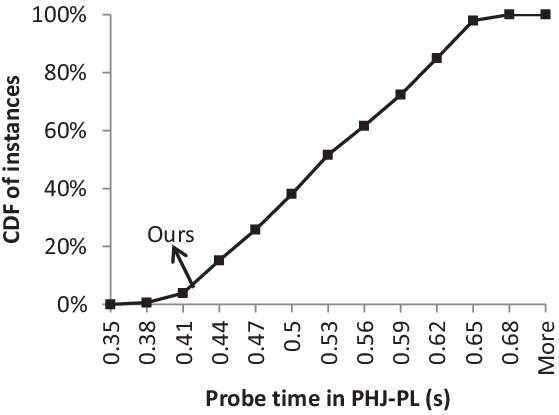}
\label{fig:cdfprobetimeinphj}}}  \vspace{-2ex}\caption{CDF of SHJ-PL
(build) and PHJ-PL (probe) with one thousand Monte Carlo
simulations} \label{fig:costModelValidation}\vspace{-2ex}
\end{figure}

\saven{To evaluate our cost model in more details, we perform
experiments with Monte Carlo simulations on the workload ratios.
Each simulation runs a PL with randomly generated ratio settings.
Figure~\ref{fig:costModelValidation} demonstrates the cumulative
distribution function (CDF) for the elapsed time of the build phase
of SHJ-PL and the probe phase of PHJ-PL with one thousand simulation
runs. We also highlight the elapsed time of our approach given by
the cost model. The proposed approach is very close to the best
performance of the Monte Carlo simulations. For each simulation run,
we evaluate the prediction by our model and the measured execution
time. The difference is smaller than $15\%$ in most cases. }

\subsection{Design Tradeoffs on Coupled CPU-GPU
Architectures}\label{subsec:tradeoff}



We evaluate the impact of each tradeoff by varying the setting of
that tradeoff while other tradeoffs have already been tuned. Based
on them, we optimize the performance of the hash join variants on
the target coupled architecture.

\begin{figure}
    \centering
    \includegraphics[width=0.32\textwidth]{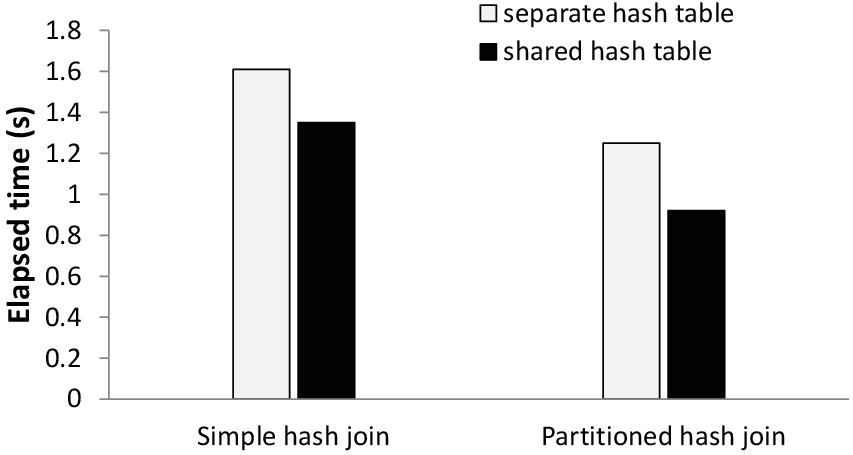}
    \vspace{-2ex}
    \caption{Elapsed time of the build phase in DD with separate and shared hash tables}\label{fig:separeteandsharedtable}
    \vspace{-2ex}
\end{figure}

\textbf{Separate vs. shared hash table.} We first show the elapsed
time of build phase of DD with separate \hejiong{and} shared hash tables, as
shown in Figure~\ref{fig:separeteandsharedtable}. SHJ-DD and PHJ-DD
with a shared hash table outperform those with separate hash tables
by 16\% and 26\%, respectively. The major benefit of adopting
the shared hash table is the elimination of the merging operation
(as we have seen in Section~\ref{subsec:discreteeval}). The other
benefit is the improved cache performance due to potential cache
reuse on the coupled architecture. The \hj{numbers} of cache misses of
SHJ-DD and PHJ-DD with separate hash tables are 2\% and 4\% larger
than those with shared hash table, respectively. The improvement
shows that, on the tested coupled architecture, the concurrency
overhead of the shared hash table is compensated by the two major
benefits over the separate hash table.



\begin{figure}
\centerline{ \subfigure[Elapsed time (sec.)
]{\includegraphics[width=0.23\textwidth]{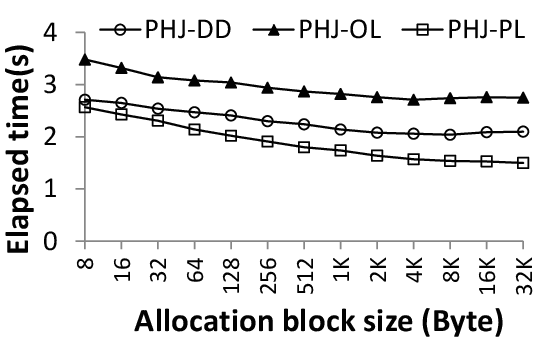}
\label{fig:dynamicmem_timePHJ}} \hfil \subfigure[Lock overhead
]{\includegraphics[width=0.23\textwidth]{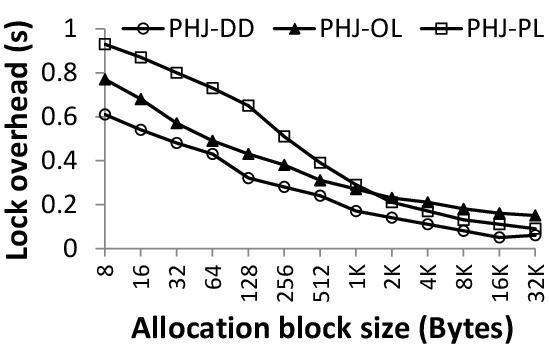}
\label{fig:dynamicmem_counterPHJ}}}  \vspace{-3ex}\caption{Elapsed
time (left) and lock overhead (right) of PHJ}
\label{fig:allocatefreememoryPHJ}\vspace{-2ex}
\end{figure}

\textbf{Optimized memory allocator.} We adopt an optimized memory
allocator to reduce the number of atomic operations.
Figure~\ref{fig:dynamicmem_timePHJ} shows the overall time with the
memory allocation block size varied in PHJ. We observed similar
results on SHJ. Overall, at the beginning, the performance keeps
\hejiong{improving} when the block size becomes larger. However, the
performance \hejiong{remains} stable when the block size is larger than 2KB. The
suitable block size is set to 2KB in our experiments. The major
reason of the performance improvement is the reduction of global
memory lock overhead from atomic operations.


To confirm our conclusion, we further study the lock overhead with
the block size varied in Figure~\ref{fig:dynamicmem_counterPHJ} for
PHJ. So far, there is no profiling tool to measure the lock
overhead directly on the hardware. Therefore, we estimate the lock
overhead as the difference of the measured time and estimated time
based on our cost model. This is because our cost model does not
consider the lock overhead. This back-of-the-envelop approach is
simplistic but sufficient to show the trend of the lock overhead for
our purpose. As the allocation memory block size increases, the lock
overhead decreases significantly for all hash join variants.

\begin{figure}
\centerline{ \subfigure[SHJ
]{\includegraphics[width=0.23\textwidth]{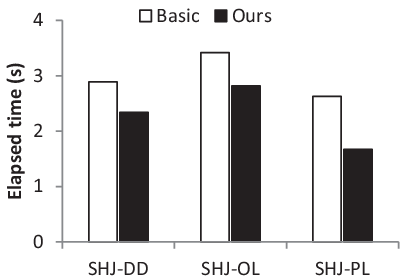}
\label{fig:allocatorinshj}} \hfil \subfigure[PHJ
]{\includegraphics[width=0.23\textwidth]{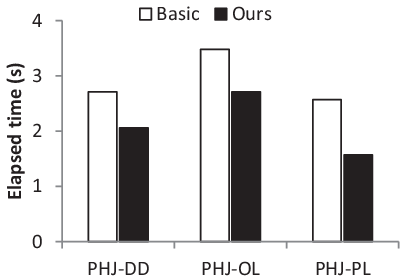}
\label{fig:allocatorinphj}}}  \vspace{-3ex}\caption{Comparison
between basic and optimized memory allocators (Basic and Ours,
respectively)} \label{fig:allocatorcomparison}\vspace{-2ex}
\end{figure}

\hejiong{Figure~\ref{fig:allocatorcomparison} compares the
performance of hash joins with the basic memory allocator and our
optimized memory allocator (denoted as Basic and Ours,
respectively). Compared with \hejiong{the} Basic allocator, the
optimized allocator significantly improves the hash join
performance, with up to 36\% and 39\% improvement on SHJ and PHJ.}

\textbf{Workload divergence.} We study the impact of grouping
approach on reducing the workload divergence. Our experimental
results show that, the grouping approach can improve the overall
performance \hj{by} 5--10\% (Figures are omitted due to space interests).
The impact on the GPU is larger than that on the CPU, because the
GPU hardware executes the wavefront strictly in a SIMD manner and
also does not have advanced branch prediction mechanisms like the
CPU.

\textbf{Fine-grained step definition.} We study the CPU and the GPU
running time for hash joins employing the fine-grained and
coarse-grained step definition. We apply PL to the coarse-grained
step definition that we \hejiong{discussed} in Section~\ref{subsec:details}, and
denote this co-processing to be PHJ-PL'. Table
\ref{table:differentstepdefinition} shows the execution time and the
cache performance of PHJ-PL and PHJ-PL'. PHJ-PL' is much slower than
PHJ-PL. As the coarse-grained step definition introduces separate
hash tables, PHJ-PL' has a larger number of cache misses and a
higher cache miss ratio than PHJ-PL. This shows the advantage of
co-processing with fine-grained steps.


\begin{table}[htb]
\begin{small}
    \centering\vspace{-2ex}
\caption{Comparison between fine-grained and coarse-grained step
definitions in PL} \label{table:differentstepdefinition}
\begin{tabular}{|p{1.3cm}|p{2.1cm}|p{2cm}|p{1.2cm}|}
\hline & L2 cache misses ($\times 10^6$) & L2 cache miss ratio &
Time (s) \\ \hline  PHJ-PL & 7 & 10\% & 1.6 \\
\hline PHJ-PL' & 15 & 23\% & 2.2 \\ \hline
\end{tabular}
\end{small}
\end{table}




\begin{figure}
\centerline{ \subfigure[SHJ
]{\includegraphics[width=0.23\textwidth]{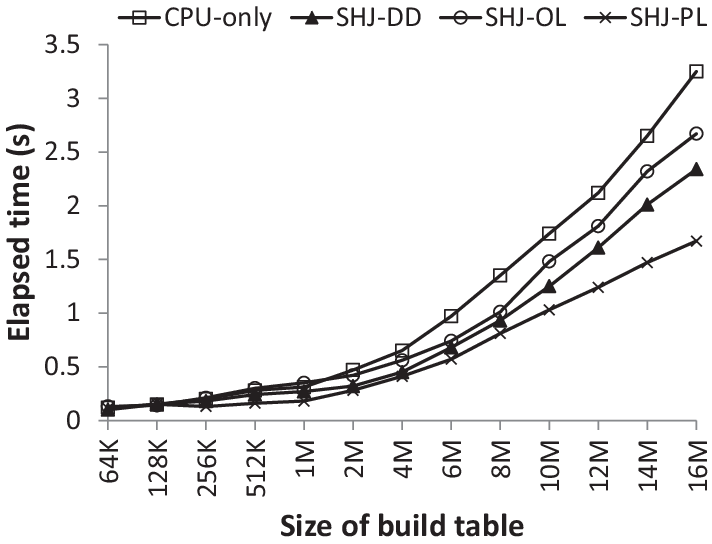}
\label{fig:endtoendcomparison1a}} \hfil \subfigure[PHJ
]{\includegraphics[width=0.23\textwidth]{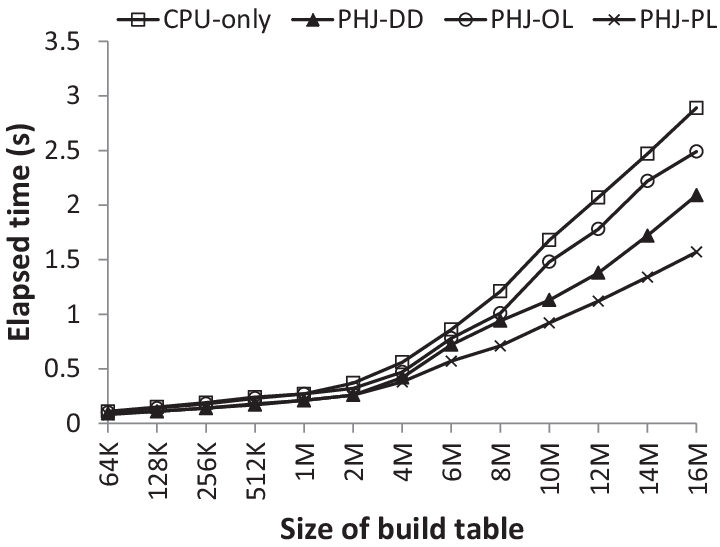}
\label{fig:endtoendcomparison1b}}} \vspace{-2ex} \caption{Elapsed
time comparison on the uniform data set}
\label{fig:endtoendcomparison1}\vspace{-2ex}
\end{figure}


\begin{figure}
\centerline{ \subfigure[SHJ
]{\includegraphics[width=0.23\textwidth]{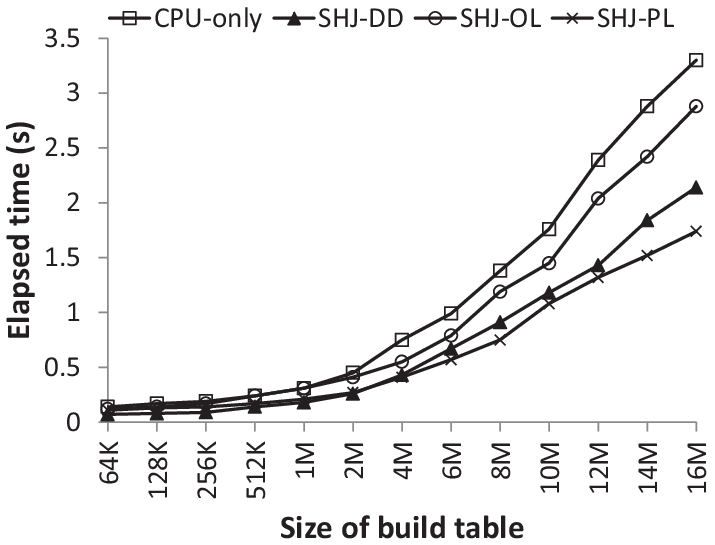}
\label{fig:endtoendcomparison3a}} \hfil \subfigure[PHJ
]{\includegraphics[width=0.23\textwidth]{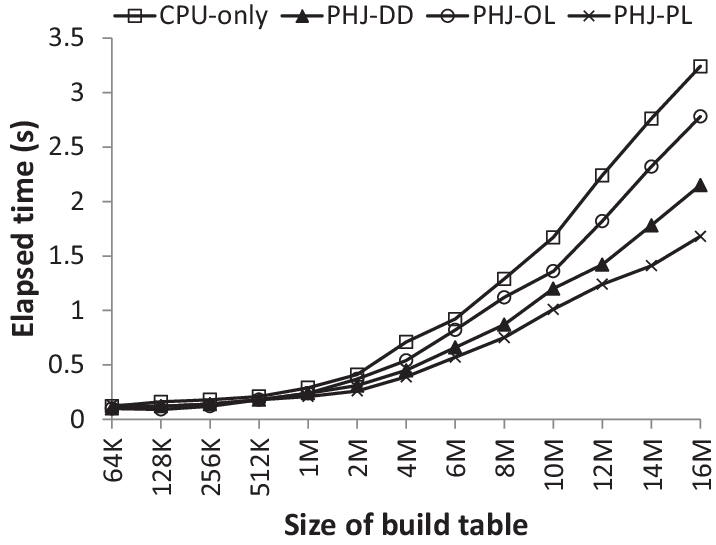}
\label{fig:endtoendcomparison3b}}} \vspace{-2ex} \caption{Elapsed
time comparison on the high-skew data set}
\label{fig:endtoendcomparison3}\vspace{-2ex}
\end{figure}

\subsection{End-to-End Performance Comparison}\label{subsec:cop}

In this section, we compare end-to-end performance for different
hash join variants on the coupled architecture. The OL essentially
is \hejiong{a} GPU-only implementation, and thus the results for GPU-only are
omitted. 

\textbf{Data skew and data size.} We first show performance numbers
on the uniform, low-skew and high-skew data sets. We fix the
probe relation to 16 million tuples and vary the build relation from
\hejiong{64K} to \hejiong{16M} tuples.
Figures~\ref{fig:endtoendcomparison1} and
\ref{fig:endtoendcomparison3} show the comparison results for
uniform and high-skew data sets, respectively. The figures for the
low-skew data sets are omitted due to the space limits, and we
describe the results in texts. In general, they have similar
performance trends on all three data sets. It shows that our
co-processing techniques work well on either uniform or skew data
sets. \saven{The elapsed time of running on high-skew data can be
actually comparable to or even lower than that on uniform data. This
is mainly because the benefit of data locality in high-skew data
compensates the overhead of latches. This point has also been
pointed out in the previous study~\cite{hashJoinOnMultiCoreCPU}. }
\hejiong{We have also observed that when the relation size exceeds
the cache size (4MB), there is a leap in the running time.}
Additionally, the performance improvement also scales well with the
build relation size. In most cases, exploiting the capabilities of
the both the GPU and the CPU (DD and Pipelined) can generate better
performance than using only the CPU (CPU-only) or GPU-only (OL).
Specifically, the optimized fine-grained pipelined join outperforms
the CPU-only and GPU-only implementations by up to 53\% and 35\%,
respectively. Thus, it is important to exploit both the CPU and the
GPU on the coupled architecture. Additionally, PL outperforms DD by
up to 28\%. This confirms the effectiveness of our optimizations on
fine-grained co-processing.




%

\begin{figure}
    \centering
    \includegraphics[width=0.32\textwidth]{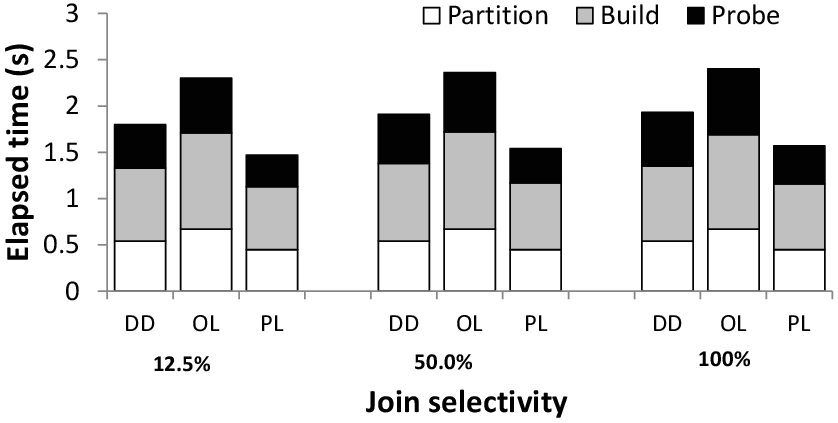}
    \vspace{-2ex}
    \caption{PHJ with join selectivity varied}\label{fig:varyjoinselectivity}
    \vspace{-2ex}
\end{figure}

\textbf{Join selectivity.} We study the impact of join selectivity.
Figure~\ref{fig:varyjoinselectivity} shows the results on time
breakdown for the join selectivity \hejiong{values of }12.5\%, 50\% and 100\%. For DD
and OL, the selectivity only affects the probe phase. As a result,
the time for the probe phase \hejiong{becomes slightly longer} when the
selectivity increases from 12.5 \% to 100\% for DD (from 0.47 to
0.58 seconds) or OL (from 0.59 to 0.71 seconds). For PL, both the
build and probe phases are affected by the join selectivity.
However, the performance impact is marginal\saven{, because our
implementation simply outputs the matching $\mathit{rid}$ pair. } When the
selectivity increases, there are more chances for workload \hejiong{divergence}
when probing the hash table. 

After studying all hash join variants, we find that PHJ-PL is
usually the fastest, which is 2--6\% faster than SHJ-PL.
Nevertheless, both SHJ-PL and PHJ-PL are very competitive on the
coupled architecture on different data sets.

\subsection{Summary and Lessons Learnt}

We summarize the major findings and lessons learnt from this study
and their implications for developing an efficient query processor
on the coupled architecture.

\hejiong{Firstly}, conventional co-processing of hash
joins~\cite{databaseOperatorsOnGPU, sortHashRevisited} can achieve
only marginal performance improvement on the coupled architecture,
since only the data transfer overhead is eliminated (accounting for
4--10\% of the total execution time). This calls for the
\hejiong{use of} fine-grained co-processing on the coupled
architecture, instead of coarse-grained co-processing in the
previous
studies~\cite{RelationalQueryCoProcessing,wenbinVLDB,databaseOperatorsOnGPU}.
Moreover, fine-grained co-processing schemes are inefficient on the
discrete architecture. It indicates \hejiong{that }``one size does
not fit all", and we need to adopt different co-processing schemes
between discrete and coupled architectures.

\hejiong{Secondly}, fine-grained co-processing on hash joins has an interesting
design space. It further increases the number of tuning knobs for
hash joins. Automaticity of optimization is feasible which is enabled by our
cost model. Going beyond hash joins, we believe that the
automaticity could be achieved through integrating the awareness of
fine-grained co-processing into query optimizations.

\hejiong{Thirdly}, memory optimizations and \hj{OpenCL-related
optimizations} have significant impact on the overall performance of
hash join co-processing. Traditional issues like shared vs. separate
hash tables remain important on the coupled architecture. More
generally, architecture-aware query processing techniques
(e.g.,~\cite{balkesen13,ross2007}) are still relevant to performance
optimizations. But their impacts should be carefully studied on the
target coupled architecture.

\hejiong{Fourthly}, fine-grained co-processing is a must on the
coupled architecture, which significantly outperforms the
performance of processing on a single processor and the conventional
co-processing. It not only keeps both processors busy, but also
assigns the suitable workload to them for efficiency. The design
space applies to general query processing, and an efficient query
processor should expose sufficiently fine-grained data parallelism
for scheduling among different processors. Additionally, we
conjecture that the design space of co-processing in this study is
relevant to not only the coupled CPU-GPU architectures but also
other heterogeneous
architectures~\cite{Kumar:2006:CAO:1152154.1152162}.

\section{Conclusion} \label{sec:conclusion}

While GPU co-processing has shown significant performance speedups
on main memory databases, the discrete CPU-GPU architecture design
has hindered the further performance improvement of GPU
co-processing. This paper revisits co-processing for hash joins on
the coupled CPU-GPU architecture. With the integration of the CPU
and the GPU into a single chip, the data transfer via PCI-e bus is
eliminated on the coupled architecture. More importantly, the
coupled architecture has opened up vast opportunities for improving
the performance of co-processing of hash joins, as we have
demonstrated in this experimental study. Specifically, we revisit
the design space of fine-grained co-processing on hash joins with
and without partitioning. We show that the fine-grained
co-processing can improve the performance by up to 53\%, 35\% and
28\% over the CPU-only, GPU-only and conventional CPU-GPU
co-processing, respectively. \saven{This paper represents a key
first step in designing efficient query co-processing on coupled
architectures.} We are developing a full-fledged query processor on
the coupled architecture~\cite{shuhaodemo}.

\section{Acknowledgement} \label{sec:acknowledgement}

\tosaven{The authors would like to thank anonymous reviewers, Qiong
Luo and Ong Zhong Liang for their valuable comments. This work is
partly supported by a MoE AcRF Tier 2 grant (MOE2012-T2-2-067) in
Singapore and an Inter-disciplinary Strategic Competitive Fund of
Nanyang Technological University 2011 for ``C3: Cloud-Assisted Green
Computing at NTU Campus''.}

\begin{scriptsize}

\bibliographystyle{abbrv}
\bibliography{hashJoinBib}

\begin{thebibliography}{10}

\bibitem{Ailamaki1999}
A.~Ailamaki, D.~J. DeWitt, M.~D. Hill, and D.~A. Wood.
\newblock Dbmss on a modern processor: Where does time go?
\newblock In {\em VLDB}, 1999.

\bibitem{AMDdevsummit}
{AMD Corp.}
\newblock http://amddevcentral.com/afds/pages/default.aspx.

\bibitem{balkesen13}
C.~Balkesen, J.~Teubner, G.~Alonso, and M.~T. Oszu.
\newblock Main-memory hash joins on multi-core cpus: tunning to the underlying
  hardware.
\newblock In {\em ICDE}, 2013.

\bibitem{hashJoinOnMultiCoreCPU}
S.~Blanas, Y.~Li, and J.~M. Patel.
\newblock Design and evaluation of main memory hash join algorithms for
  multi-core cpus.
\newblock In {\em SIGMOD}, pages 37--48, 2011.

\bibitem{databaseNewBottleneck}
P.~A. Boncz, S.~Manegold, and M.~L. Kersten.
\newblock Database architecture optimized for the new bottleneck: Memory
  access.
\newblock In {\em VLDB}, 1999.

\bibitem{apumapreduce}
L.~Chen, X.~Huo, and G.~Agrawal.
\newblock Accelerating mapreduce on a coupled cpu-gpu architecture.
\newblock In {\em SC}, pages 25:1--25:11, 2012.

\bibitem{Chen:1992:USR:645918.672489}
M.-S. Chen, M.-L. Lo, P.~S. Yu, and H.~C. Young.
\newblock Using segmented right-deep trees for the execution of pipelined hash
  joins.
\newblock In {\em VLDB}, pages 15--26, 1992.

\bibitem{hashJoinPrefetching}
S.~Chen, A.~Ailamaki, P.~B. Gibbons, and T.~C. Mowry.
\newblock Improving hash join performance through prefetching.
\newblock {\em ACM Trans. Database Syst.}, 2007.

\bibitem{DeWitt:1992:PDS:129888.129894}
D.~DeWitt and J.~Gray.
\newblock Parallel database systems: the future of high performance database
  systems.
\newblock {\em Commun. ACM}, 1992.

\bibitem{DeWitt:1985:MHJ:1286760.1286774}
D.~J. DeWitt and R.~H. Gerber.
\newblock Multiprocessor hash-based join algorithms.
\newblock In {\em VLDB}, pages 151--164, 1985.

\bibitem{cudaopenclcomparison}
J.~Fang, A.~L. Varbanescu, and H.~Sips.
\newblock A comprehensive performance comparison of cuda and opencl.
\newblock In {\em ICPP}, 2011.

\bibitem{wenbinVLDB}
W.~Fang, B.~He, and Q.~Luo.
\newblock Database compression on graphics processors.
\newblock {\em Proc. VLDB Endow.}, pages 670--680, 2010.

\bibitem{pipelinedHashJoinOnMultithreadedArchi}
P.~Garcia and H.~F. Korth.
\newblock Pipelined hash-join on multithreaded architectures.
\newblock In {\em DaMoN}, pages 1:1--1:8, 2007.

\bibitem{Gold:2005:ADO:1114252.1114260}
B.~Gold, A.~Ailamaki, L.~Huston, and B.~Falsafi.
\newblock Accelerating database operators using a network processor.
\newblock In {\em DaMoN}, 2005.

\bibitem{RelationalQueryCoProcessing}
B.~He, M.~Lu, K.~Yang, R.~Fang, N.~K. Govindaraju, Q.~Luo, and P.~V. Sander.
\newblock Relational query coprocessing on graphics processors.
\newblock {\em ACM TODS}, pages 21:1--21:39, 2009.

\bibitem{hetodscacheoblivious}
B.~He and Q.~Luo.
\newblock Cache-oblivious databases: Limitations and opportunities.
\newblock {\em ACM Trans. Database Syst.}, pages 8:1--8:42, 2008.

\bibitem{databaseOperatorsOnGPU}
B.~He, K.~Yang, R.~Fang, M.~Lu, N.~Govindaraju, Q.~Luo, and P.~Sander.
\newblock Relational joins on graphics processors.
\newblock In {\em SIGMOD}, pages 511--524, 2008.

\bibitem{He:2011:HTE:1952376.1952381}
B.~He and J.~X. Yu.
\newblock High-throughput transaction executions on graphics processors.
\newblock {\em Proc. VLDB Endow.}, pages 314--325, 2011.

\bibitem{keyvalue}
T.~H. Hetherington and {et al}.
\newblock Characterizing and evaluating a key-value store application on
  heterogeneous cpu-gpu systems.
\newblock In {\em ISPASS}, pages 88--98, 2012.

\bibitem{Hong:2010:MWP:1854273.1854303}
C.~Hong, D.~Chen, W.~Chen, W.~Zheng, and H.~Lin.
\newblock Mapcg: writing parallel program portable between cpu and gpu.
\newblock In {\em PACT}, pages 217--226, 2010.

\bibitem{GPUJoinRevisited}
T.~Kaldewey, G.~Lohman, R.~Mueller, and P.~Volk.
\newblock Gpu join processing revisited.
\newblock In {\em DaMoN}, pages 55--62, 2012.

\bibitem{sortHashRevisited}
C.~Kim and {et al}.
\newblock Sort vs. hash revisited: fast join implementation on modern
  multi-core cpus.
\newblock {\em PVLDB}, 2009.

\bibitem{Kumar:2006:CAO:1152154.1152162}
R.~Kumar, D.~M. Tullsen, and N.~P. Jouppi.
\newblock Core architecture optimization for heterogeneous chip
  multiprocessors.
\newblock In {\em PACT}, pages 23--32, 2006.

\bibitem{YinanCIDR}
Y.~Li, I.~Pandis, R.~Mueller, V.~Raman, and G.~Lohman.
\newblock Numa-aware algorithms: the case of data shuffling.
\newblock In {\em CIDR}, 2013.

\bibitem{Lo:1993:OPA:170035.170053}
M.-L. Lo, M.-S.~S. Chen, C.~V. Ravishankar, and P.~S. Yu.
\newblock On optimal processor allocation to support pipelined hash joins.
\newblock In {\em SIGMOD}, pages 69--78, 1993.

\bibitem{Manegold:2002:GDC:1287369.1287387}
S.~Manegold, P.~Boncz, and M.~L. Kersten.
\newblock Generic database cost models for hierarchical memory systems.
\newblock In {\em VLDB}, pages 191--202, 2002.

\bibitem{Manegold2000}
S.~Manegold, P.~A. Boncz, and M.~L. Kersten.
\newblock What happens during a join? dissecting cpu and memory optimization
  effects.
\newblock In {\em VLDB}, 2000.

\bibitem{hashjoinOnCGwithPCIeconstraint}
H.~Pirk, S.~Manegold, and M.~Kersten.
\newblock Accelerating foreign-key joins using asymmetric memory channels.
\newblock In {\em ADMS}, 2011.

\bibitem{ross2007}
K.~Ross.
\newblock Efficient hash probes on modern processors.
\newblock In {\em ICDE}, pages 1297--1301, 2007.

\bibitem{Schneider:1989:PEF:67544.66937}
D.~A. Schneider and D.~J. DeWitt.
\newblock A performance evaluation of four parallel join algorithms in a
  shared-nothing multiprocessor environment.
\newblock In {\em SIGMOD}, pages 110--121, 1989.

\bibitem{CacheConsciousHashJoin}
A.~Shatdal, C.~Kant, and J.~F. Naughton.
\newblock Cache conscious algorithms for relational query processing.
\newblock In {\em VLDB}, 1994.

\bibitem{cstore}
M.~Stonebraker and {et al}.
\newblock C-store: a column-oriented dbms.
\newblock In {\em VLDB}, pages 553--564, 2005.

\bibitem{GPGPU:DGEMMOnGPU}
G.~Tan, L.~Li, S.~Triechle, E.~Phillips, Y.~Bao, and N.~Sun.
\newblock Fast implementation of dgemm on fermi gpu.
\newblock In {\em SC}, 2011.

\bibitem{Teubner:2011:SPS:1989323.1989389}
J.~Teubner and R.~Mueller.
\newblock How soccer players would do stream joins.
\newblock In {\em SIGMOD}, pages 625--636, 2011.

\bibitem{openmpopenclcomparison}
K.~Thouti and S.R.Sathe.
\newblock Comparison of openmp and opencl parallel processing technologies.
\newblock {\em IJACSA}, pages 56--161, 2012.

\bibitem{shuhaodemo}
S.~Zhang, J.~He, B.~He, and M.~Lu.
\newblock Omnidb: Towards portable and efficient query processing on parallel
  cpu/gpu architectures.
\newblock In {\em PVLDB}, pages 1--4, 2013.

\end{thebibliography}

\end{scriptsize}

\appendix


\section{More Experimental Results}

\saven{We present more experimental results, including comparison
with a coarse-grained scheduling approach, evaluating hash join
performance for data sets larger than zero copy buffer and more
detailed studies on latches.}



\begin{figure}[!htb]
    \centering
    \includegraphics[width=0.35\textwidth]{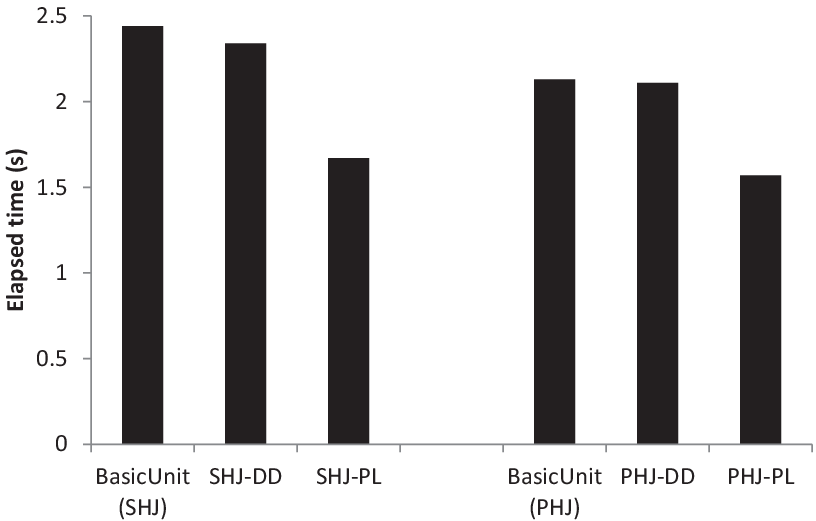}
    \vspace{-2ex}
    \caption{Performance comparison between {BasicUnit} and fine-grained co-processing algorithms}\label{fig:basicunit}
    \vspace{-2ex}
\end{figure}

\textbf{Comparison with A Coarse-Grained Scheduling Approach.} For
comparison, we have implemented a simple coarse-grained scheduling
approach (namely ``BasicUnit"). The BasicUnit approach dynamically
assigns a chunk of tuples to the CPU or the GPU, and performs the
build or probe operations with that chunk on the corresponding
processor. Due to the architectural difference between the CPU and
the GPU, the chunk size is tuned for the target architecture.
Compared with our proposed approach, the BasicUnit approach has a
major deficiency: the CPU may perform more non-CPU-optimized tasks
than our proposed fine-grained co-processing, and vice versa. We
have implemented the suggested approach, and report the results in
Figure~\ref{fig:basicunit}. When both R and S tables have 16M
tuples, our proposed approaches (SHJ-PL and PHJ-PL) are 31\% and
25\% faster than their counterparts using the BasicUnit approach.
Besides, although BasicUnit works like DD, the scheduling overhead
of BasicUnit \hj{results in performance degradation}.

\begin{figure}[!htb]
\centerline{ \subfigure[Build
]{\includegraphics[width=0.15\textwidth]{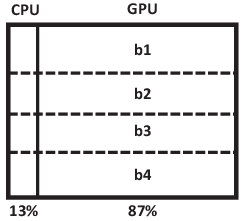}
\label{fig:ratioinbuildshj}} \hfil \subfigure[Probe
]{\includegraphics[width=0.15\textwidth]{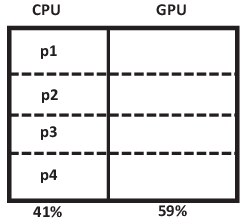}
\label{fig:ratioinprobeshj}}}  \vspace{-2ex}\caption{Workload ratios of different steps for SHJ employing BasicUnit}
\label{fig:ratioinbasicunitshj}\vspace{-2ex}
\end{figure}

\begin{figure}[!htb]
\centerline{ \subfigure[Partition
]{\includegraphics[width=0.15\textwidth]{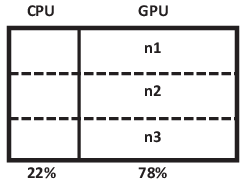}
\label{fig:ratioinpartitionphj}} \hfil \subfigure[Build
]{\includegraphics[width=0.15\textwidth]{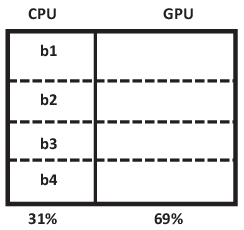}
\label{fig:ratioinbuildphj}}     \hfil \subfigure[Probe
]{\includegraphics[width=0.15\textwidth]{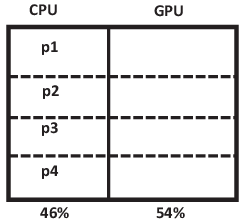}
\label{fig:ratioinprobephj}}}    \vspace{-2ex}\caption{Workload ratios of different steps for PHJ employing BasicUnit}
\label{fig:ratioinbasicunitphj}  \vspace{-2ex}
\end{figure}

We have further investigated the workload ratios of different steps
on the CPU and the GPU. Figures~\ref{fig:ratioinbasicunitshj} and
\ref{fig:ratioinbasicunitphj} show the results using the BasicUnit
approach for SHJ and PHJ, respectively. The ratios are significantly
different from the optimal ratios given by our cost model (Figures
\ref{fig:SHJratio} and \ref{fig:PHJratio} in the paper) because the
CPU or the GPU has to process the same portion of the workload in
all steps of a given phase, resulting in deficiency compared with
our fine-grained co-processing.



\textbf{Large datasets that cannot fit into the zero copy buffer.}
In the current system, the zero copy buffer size is limited. For
larger data sets that cannot fit into the zero copy buffer, we
perform the hash join with partitioning. The algorithm is similar to
classic hash joins for external memory. Here, we view the zero copy
buffer as main memory and other system memory as external memory.
Specifically, we partition the build and the probe relations so that
the join of a partition pair can fit into zero copy buffer. The
partitioning can have multiple passes. In partition phase, a block
of tuples from the input relation is partitioned in the zero copy
buffer with our current partitioning algorithm. In our experiment,
the chunk size is 16 million tuples. Next, we copy the intermediate
partitions out from the zero copy buffer to the system memory. After
all the inputs are partitioned, we link all the intermediate
partitions together \hj{to} form the result partition pairs. Next,
we apply the hash join algorithms proposed in this paper to handle
the join in each partition pair.

\begin{figure}[!htb]
    \centering
    \includegraphics[width=0.40\textwidth]{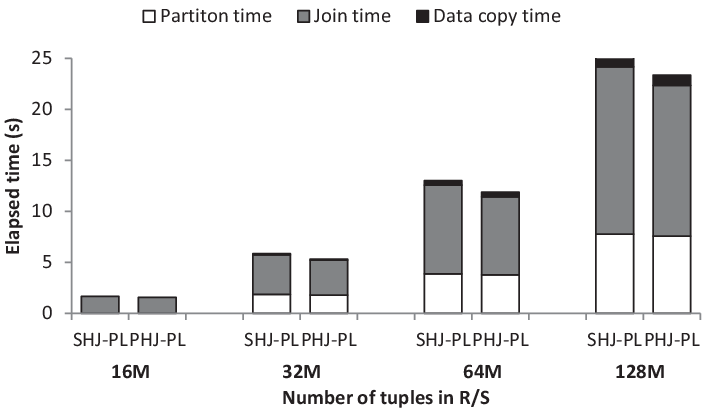}
    \vspace{-2ex}
    \caption{Comparing elapsed time of different join algorithms on each partition pair for large data sets ($|R|=|S|$)}\label{fig:largesizehj}
    \vspace{-2ex}
\end{figure}

Figure~\ref{fig:largesizehj} shows the performance when we vary the
number of tuples in both relations $R$ and $S$ from 16M (256MB) to
128M (2GB). We compare different join algorithms on the partition
pair, either SHJ-PL or PHJ-PL. PHJ-PL is \hejiong{up to 9\%} faster
than SHJ-PL. We divide the elapsed time into \hj{three} parts: data
copy time, \hj{partition time and} join time for the time on data
transfer between the system memory and zero copy buffer, \hj{the
time for partitioning and} for performing the join on the partition
pairs, respectively. There is no data copy time or partition time
when $|R|=|S|=16M$, because the entire join can fit into the zero
copy buffer. On other cases, partition time is very significant. We
obtain the data copy time from AMD CodeAnalyst profiler. The data
copy time is \hejiong{around 4\%} of the total time.
\hj{Furthermore, both the partition time and the join time increase
almost linearly with the size of input tables. It indicates good
scalability of our algorithm when the data size is larger than zero
copy buffer.}

%

\begin{figure}[!htb]
\centerline{ \subfigure[Locking time on the CPU
]{\includegraphics[width=0.23\textwidth]{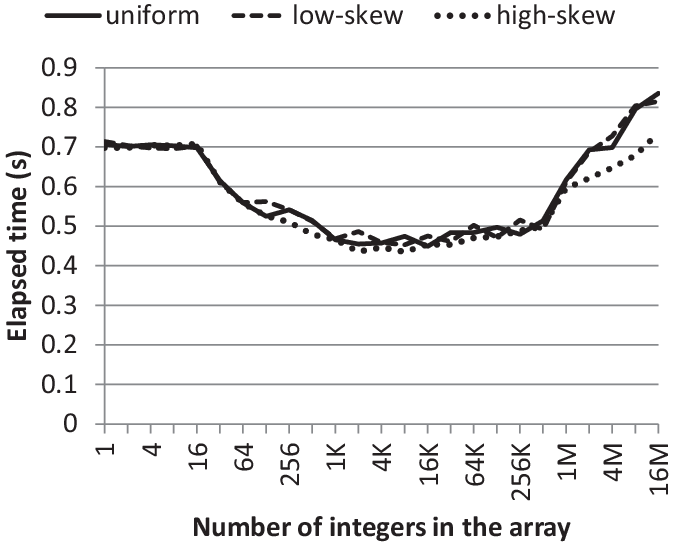}
\label{fig:lockingoncpu}} \hfil \subfigure[Locking time on the GPU
]{\includegraphics[width=0.23\textwidth]{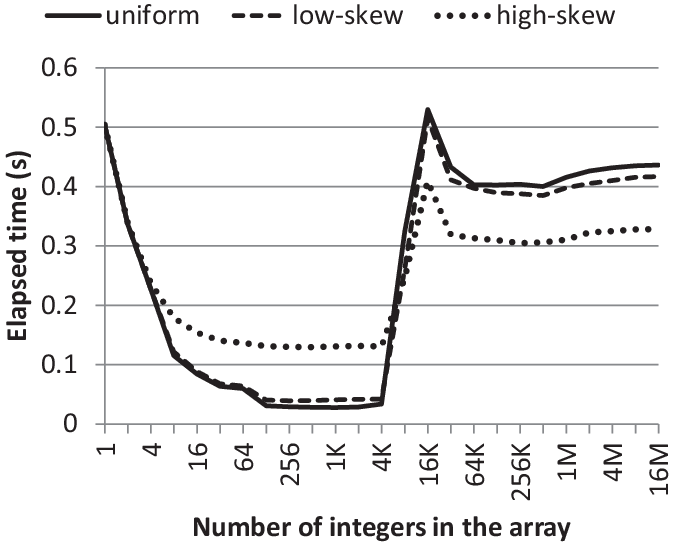}
\label{fig:lockingongpu}}}  \vspace{-2ex}\caption{The performance of
micro benchmark on locking overhead on the CPU and the GPU}
\label{fig:lockingoverhead}\vspace{-2ex}
\end{figure}

\textbf{Locking overhead.} We have designed a micro benchmark to
study the performance impact of our latch implementation on APU. In
the micro benchmark, we first generated an array of $N$ integers and
then we used $K$ threads to perform $X$ increments in total on the
array, where $X$ is a constant. The increment \hejiong{operation} is
implemented with our latch. We investigated different data
distributions (the notations of ``uniform'', ``low-skew'' and
``high-skew'' have the same meanings as the experimental setting in
the paper). The results are shown in
Figure~\ref{fig:lockingoverhead}, where $N$ ranges from 1 to 16M,
$X$ is fixed to be 16M on both the CPU and the GPU, $K$ is fixed to
be 8192 and 256 on the GPU and the CPU respectively.  For both the
CPU and the GPU, the overhead decreases when the data size increases
until the input array cannot fit into the cache size {(4MB)}. After
that, the execution time of running on the high-skew data is
slightly lower than that on the uniform data. This is mainly because
of the benefit of data locality in the high-skew data that
compensates the overhead of latches. This point has also been
pointed out in Section 4.4 of the previous study
\cite{hashJoinOnMultiCoreCPU}.

\end{document}